\documentclass[aps,prx,letterpaper,twocolumn,floats,superscriptaddress,tighten,eqsecnum]{revtex4}

\UseRawInputEncoding

\usepackage{amssymb}
\usepackage{amsbsy}
\usepackage{amsmath}
\usepackage{graphicx}
\usepackage{graphics}
\usepackage{setspace}
\usepackage{array}
\usepackage{color}
\usepackage{xcolor}
\usepackage{fontenc}
\usepackage{textcomp}
\usepackage{rotating}
\usepackage{bm}
\usepackage{hyperref}
\usepackage{placeins}
\hypersetup{pdfpagemode=UseNone}
\usepackage{subfigure}
\usepackage{chngcntr}

\newcommand{\e}{\varepsilon}

\newcommand{\vex}[1]{\bm{\mathrm{#1}}}

\newcommand{\bsub}{\begin{subequations}}
\newcommand{\esub}{\end{subequations}}

\newcommand{\be}{\begin{equation}}
\newcommand{\ee}{\end{equation}}
\newcommand{\bea}{\begin{eqnarray}}
\newcommand{\eea}{\end{eqnarray}}

\begin{document}
\title{Acoustic phonon contribution to the resistivity of twisted bilayer graphene}
\author{Seth M.\ Davis}
\email{smdavis1@umd.edu}
\affiliation{Condensed Matter Theory Center and Joint Quantum Institute, Department of Physics, University of Maryland, College Park, MD 20742, USA}
\author{Fengcheng\ Wu}
\affiliation{School of Physics and Technology, Wuhan University, Wuhan 430072, China}
\affiliation{Wuhan Institute of Quantum Technology, Wuhan 430206}
\author{Sankar\ Das\ Sarma}
\affiliation{Condensed Matter Theory Center and Joint Quantum Institute, Department of Physics, University of Maryland, College Park, MD 20742, USA}
\date{\today}

\begin{abstract}
We calculate the contribution to the doping ($n$) and temperature ($T$) dependence of the electrical resistivity of twisted bilayer graphene (TBLG) due to scattering by acoustic phonons. Our calculation retains the full Bistritzer-MacDonald (BM) band structure, with a focus on understanding the role of the complicated geometric features present in the BM band structure on electronic transport theory. We find that the band geometry plays an important role in determining the resistivity, giving an intricate dependence on both $n$ and $T$ that mirrors features in the band structure and complicates the Bloch-Gr\"{u}neisen (BG) crossover. Our calculations predict pronounced departures from the standard simplistic expectation of a linear $T$-dependence above the BG crossover. In particular, we are able to explain the presence of the resistance peaks that have been observed in experiment, as well as quantitatively predict the temperatures at which they occur. Our calculated theoretical results are germane to an ongoing debate over the existence of a strange metal state in TBLG by providing a quantitatively accurate theory for the TBLG resistivity at finite temperatures.
\end{abstract}
\maketitle




\section{Introduction}
\label{Section-Introduction}

\begin{figure}[t!]
\includegraphics[angle=0,width=.47\textwidth]{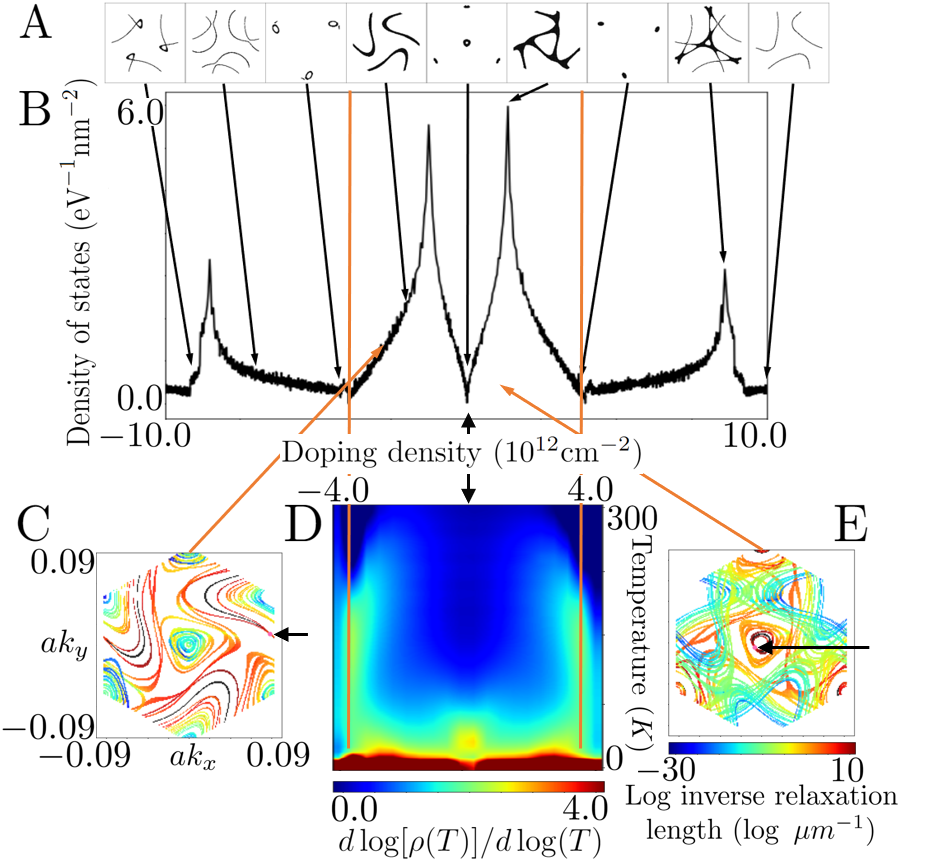}
\caption{We depict the complex band geometry at play in twisted bilayer graphene scattering processes. Panel (B) plots the density of states of TBLG for the twist angle $\theta = 1.3^{\circ}$, while (A) depicts the Fermi surface geometry at various doping levels. Orange bars denote the edges of the nearly flat bands. Panel (D) gives an power law for the resistivity, calculated via $\partial \log[\rho(n,T)]/\partial \log T$. Comparison of (D) and (B) shows the extent to which the $T$-dependence of resistivity is altered due to the geometry of the band structures. Panels (C) and (E) depict \textit{scattering manifolds} for TBLG, showing the set of kinematically allowed scattering states for a given reference state marked by arrows. The log of the individual scattering rates between Bloch states are given by the color plot. The calculation is done at $100K$ and for chemical potentials fixed at $\mu = -0.015\ eV$ (C) and $\mu = 0.005\ eV$ (E).}
\label{IntroductionDoSFigure}
\end{figure}

Following the discovery of superconductivity (SC) proximate to correlated insulator states in magic angle twisted bilayer graphene (TBLG), \cite{Cao_2018a, Cao_2018b, Cao2020PRL, Cao2021, Yankowitz_2019, Kerelsky_2019, Lu_2019, Wu_2018} understanding the nature of SC in TBLG has become a central contemporary challenge in condensed matter physics. While the SC transition critical temperature ($T_c$) is not particularly high ($T_c \approx 3K$), the presence of the flat bands produced at the magic angle increase the relative importance of electron-electron interaction and render the problem ``strongly correlating" \cite{Bistritzer_2011}. The fact that a 2D strongly correlated system displays phenomenology similar to that seen in the high-$T_c$ superconductors (e.g. neighboring insulating phases, strong doping dependence) has inspired hopeful speculation that the SC in magic angle TBLG has its origin in the same strong correlation physics. If so, further experiment on TBLG could offer clues that could finally lead to the solution of the long-standing high-$T_c$ SC problem. But, in sharp contrast to the high-$T_c$ cuprates, there is also strong experimental evidence favoring a physical picture for TBLG in which the SC exists at all dopings except for at commensurate moir\'{e} filling fractions, where it is preempted due to strong correlation effects. This picture may imply that SC and strongly correlated insulating phases actually compete in TBLG, and arise from completely different underlying mechanisms, such as electron-phonon interactions in the case of the SC (as in most SC materials) and electron-electron interactions in the case of the insulators \cite{Lu_2019}.

\begin{figure*}[t]
\includegraphics*[angle=0,width=.9\textwidth]{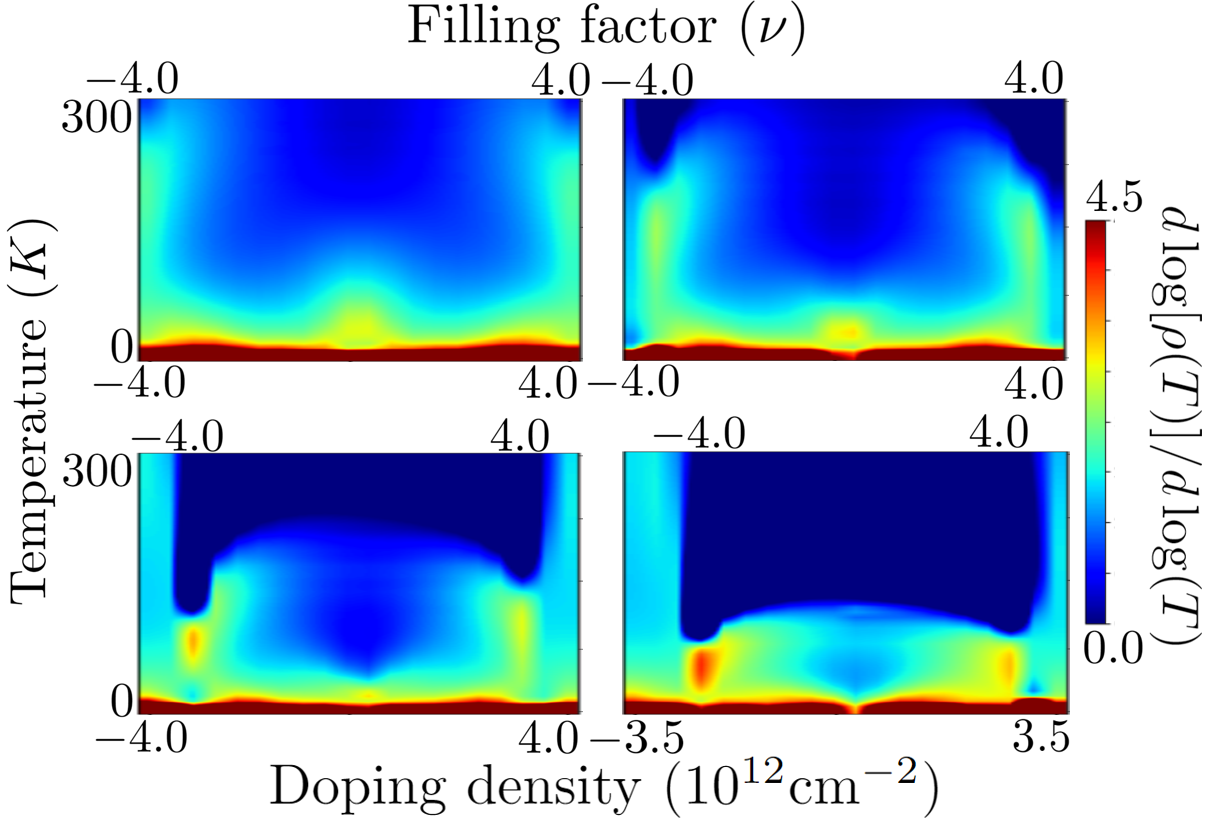}
\caption{We give approximate scaling power-laws for the $T$-dependence of the resistivity as a function of both doping ($n$) and temperature ($T$), extracted via $\partial \log[\rho(n,T)]/\partial \log T$. We present results for several different twist angles: $\theta = 1.4^{\circ}$ (top left), $\theta = 1.3^{\circ}$ (top right), $\theta = 1.2^{\circ}$ (bottom left), $\theta = 1.1^{\circ}$ (bottom right). These graphs depict the Bloch-Gr\"{u}neisen crossover, which predicts a $T^4$ scaling for resistivity at low $T$ and a linear-in-$T$ scaling at high $T$, above a crossover temperature $T^*_{BG}$ [cf. Eq.~(\ref{BGTransitionTemperature})]. We emphasize that our results predict large regions where the resistivity scales sub-linearly with $T$, depicted in dark blue. These $T$-nonlinear regions are due to the band geometry. Near the magic angle, the band structure geometry is quite sensitive to twist angle. In these results, we can see how this sensitivity is passed on to the $T$-dependence of resistivity. The bottom of each plot marks the filling in terms of doping density, while the top of each plot gives the filling in terms of the filling factor [see Eq.~(\ref{FillingFactor})], marking with $\pm4.0$ the ends of the first moir\'{e} valence and conduction bands.} 
\label{TScalingPhaseDiagramFigure}
\end{figure*}

\begin{figure*}[t]
\includegraphics*[angle=0,width=.9\textwidth]{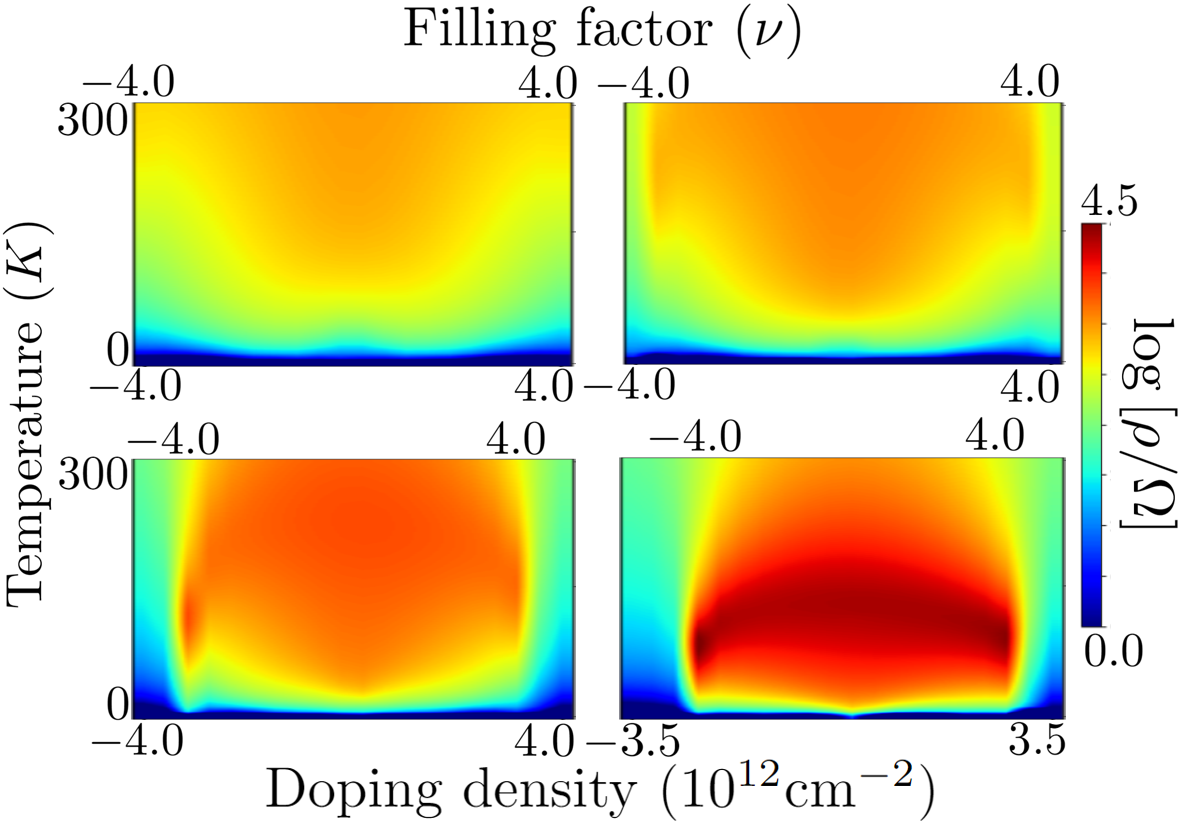}
\caption{We present the ($log$) resistivity as a function of both doping ($n$) and temperature ($T$) for several different twist angles: $\theta = 1.4^{\circ}$ (top left), $\theta = 1.3^{\circ}$ (top right), $\theta = 1.2^{\circ}$ (bottom left), $\theta = 1.1^{\circ}$ (bottom right). These plots give a global view of our resistivity predictions in $(n,T)$ space. We emphasize that the band curvature produces peaks in the resistivity near charge-neutral doping in an intermediate temperature range of $30K-200K$. The bottom of each plot marks the filling in terms of doping density, while the top of each plot gives the filling in terms of the filling factor [see Eq.~(\ref{FillingFactor})], marking with $\pm4.0$ the ends of the first moir\'{e} valence and conduction bands.}
\label{LogRhoTFigure}
\end{figure*}

\begin{figure*}[t!]
\includegraphics[angle=0,width=0.48\textwidth]{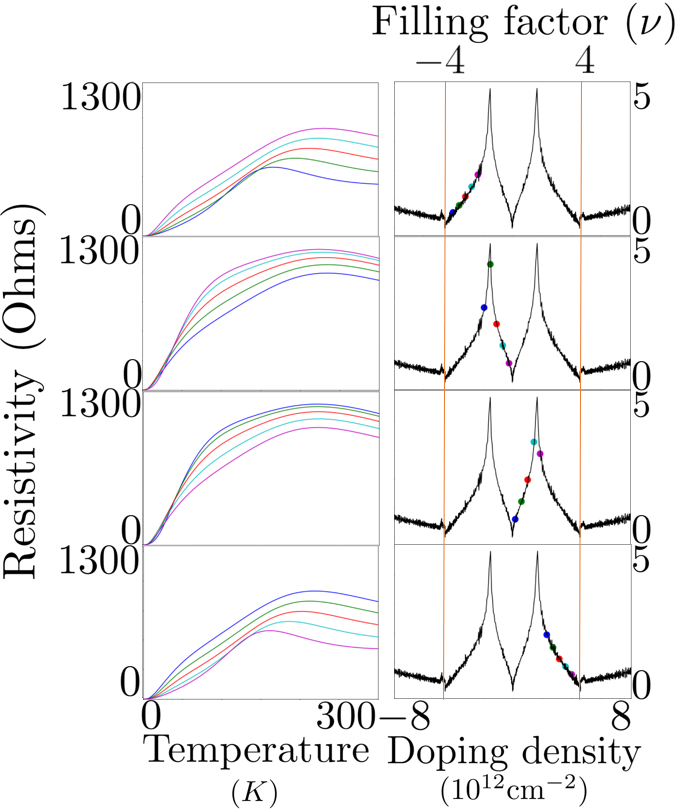}
\includegraphics[angle=0,width=0.48\textwidth]{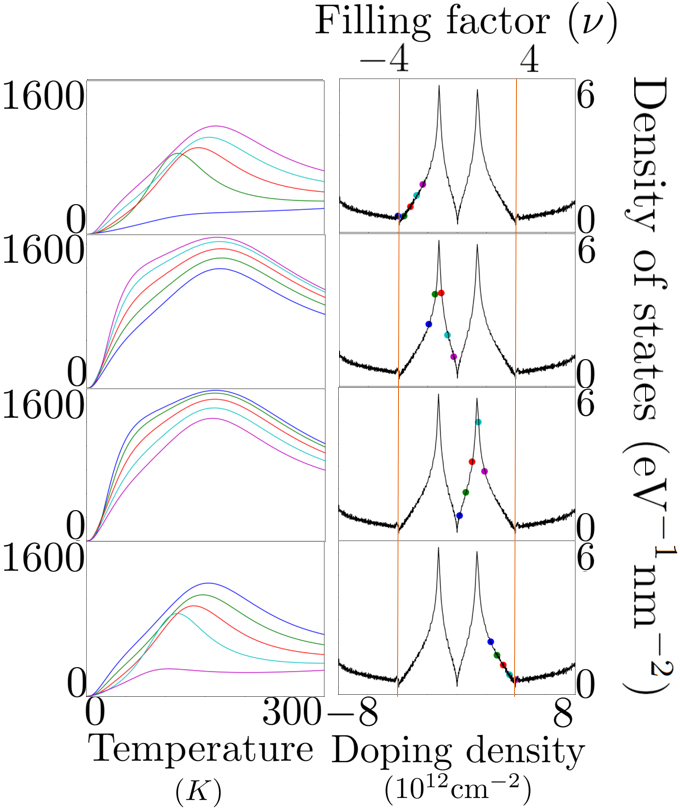}
\caption{Curves depicting the temperature ($T$) dependence of the resistivity [$\rho(n,T)$] for various dopings ($n$). The left portion of the figure is for TBLG at twist angle $\theta = 1.4^{\circ}$, while the right portion is for $\theta = 1.3^{\circ}$. We plot resistivity over the temperature range [$0K,300K$], and we denote on the density of states plot which dopings the $\rho(T)$ curves correspond to. Each color of the resistivity plot corresponds to the doping level shown in the discrete color points in the corresponding density of states curves next to it. In the density of states plots, orange lines mark the $\nu = \pm 4$ fillings [see Eq.~(\ref{FillingFactor})], which correspond to the edges of the first moir\'{e} valence and conduction band.}
\label{PrimarySlicesFigure}
\end{figure*}

\begin{figure*}[t!]
\includegraphics[angle=0,width=0.48\textwidth]{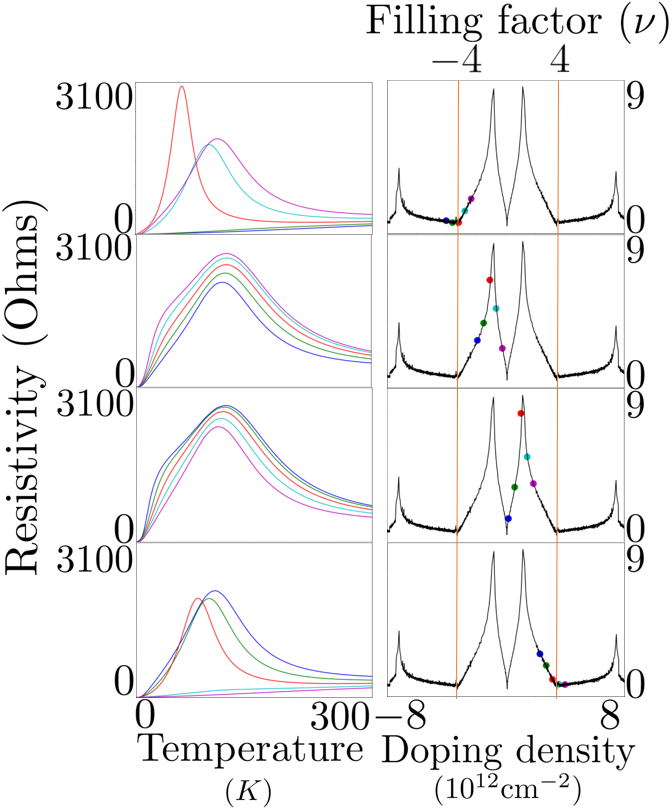}
\includegraphics[angle=0,width=0.50\textwidth]{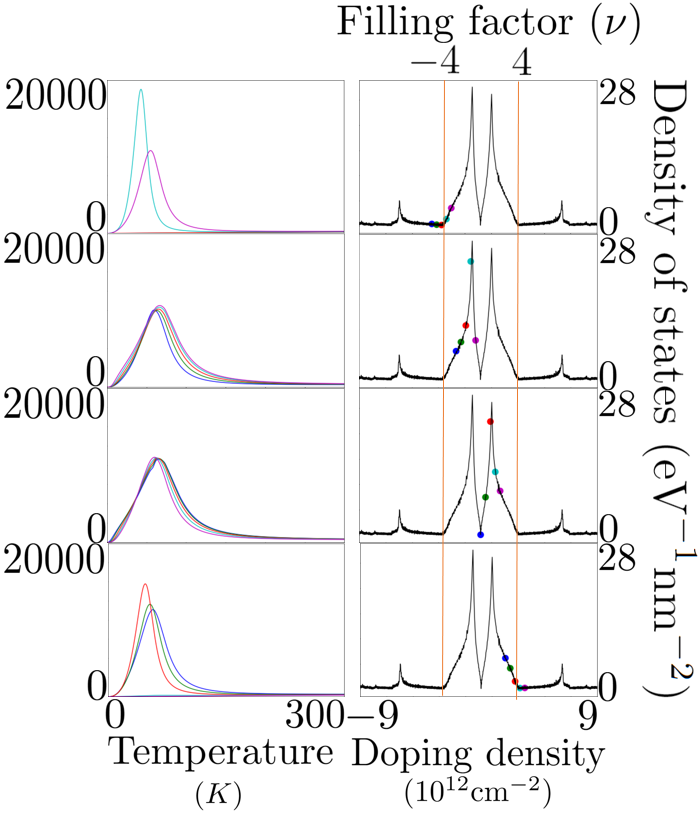}
\caption{Curves depicting the temperature ($T$) dependence of the resistivity [$\rho(n,T)$] for various dopings ($n$). The left portion of the figure is for TBLG at twist angle $\theta = 1.2^{\circ}$, while the right portion is for $\theta = 1.1^{\circ}$. We plot resistivity over the temperature range [$0K,300K$], and we denote on the density of states plot which dopings the $\rho(T)$ curves correspond to. Each color of the resistivity plot corresponds to the doping level shown in the discrete color points in the corresponding density of states curves next to it. In the density of states plots, orange lines mark the $\nu = \pm 4$ fillings [see Eq.~(\ref{FillingFactor})], which correspond to the edges of the first moir\'{e} valence and conduction band.}
\label{SecondarySlicesFigure}
\end{figure*}

Motivated by the above, there has been a huge amount of theoretical work on TBLG over the last few years, investigating its properties and drawing various connections to other strongly correlated systems capable of SC. On the other hand, theories of phonon-mediated BCS-style SC have also been put forth that are able to predict roughly-accurate  transition temperatures for TBLG SC \cite{Wu2019_phonon, Li2020}. In addition, there is direct experimental support for electron-phonon interaction induced SC in TBLG: external gating enhances (suppresses) the SC phase (the correlated insulating phase), presumably because gating reduces the electron-electron interaction through screening, leading to stronger SC by virtue of the suppression of the repulsive Coulomb interaction \cite{Stepanov2020untying, Saito2020independent}. Further, the effective electron-phonon coupling in TBLG - estimated based on the $T_c$ for SC using the standard BCS theory - agrees with that estimated from (phonon-limited) transport properties in TBLG, again strongly suggesting the crucial role of electron-phonon interaction in both the observed SC as well as the metallic resistivity of TBLG \cite{Wu2019_phonon, Das_Sarma_2020}.

The investigation of TBLG SC has evolved naturally into a debate as to whether or not TBLG exhibits a \textit{strange metal} phase \cite{TBLGStrangeMetalExperiment1, Polshyn2019, TBLGStrangeMetalExperiment3, SankarFengchengStrangeMetal, Das_Sarma_2020}. The strange metal - often discussed for $T > T_c$ in the cuprates and in other strongly correlated superconducting systems - is characterized by a linear-in-$T$ resistivity that spans an unusually large range of temperatures and often, but not always, has an anomalously large temperature coefficient for the resistivity \cite{Hwang_2019}. Indeed, TBLG has been reported to exhibit linear-in-T resistivity over a large range of temperatures and dopings \cite{TBLGStrangeMetalExperiment1}. However, the debate is complicated by the fact that phonon scattering \textit{also} generally produces linear-in-$T$ resistivity above a crossover temperature $T^*_{BG}$ \cite{AshcroftAndMermin, SankarGrapheneKT1, Hwang_2019}. Especially since acoustic phonons provide a plausible theory of SC in TBLG (as well as in moir\'{e}less crystalline layered graphene systems) it is important to carefully differentiate whether the $T$-linear resistivity could arise from phonon scattering. This is particularly relevant since the corresponding linear-in-$T$ resistivities of regular monolayer and bilayer graphene are well accounted for by acoustic phonon scattering with a quantitative agreement between theory \cite{Hwang2008, Min2011} and experiment \cite{Efetov_2010}.

The usual physics of acoustic-phonon-limited resistivity in metals and semimetals is as follows \cite{SankarGrapheneKT1, SankarGrapheneKT2, Wu2019_phonon, Li2020, SankarGrapheneKT5, Ziman, AshcroftAndMermin}. At low $T$, where the bosonic quantum statistics of the phonon dominate, the resistivity is characterized by a power-law scaling $\rho \propto T^{d+2}$, where $d$ is the dimension of the sample. This behavior defines the ``Bloch-Gr\"{u}neisen" (BG) regime. At higher temperatures, in the so-called \textit{equipartition regime} where the phonon thermal occupancy is basically classical,  there is then a crossover to linear scaling ($\rho \propto T$) which takes place roughly around the \textit{BG crossover temperature}, $T^*_{BG}$. In the case of a circular Fermi surface and quasi-elastic scattering ($v_p << v_F$) we find that
\begin{align}
\label{BGTransitionTemperature}
    k_BT^*_{BG} = \mathcal{C}_{BG}\cdot(2\hbar v_p k_F),
\end{align}
where $v_p$ is the phonon velocity, $k_F$ is the Fermi momentum, and $\mathcal{C}_{BG} \approx \mathcal{O}(1)$ is a material-specific constant. (Further, we define the \textit{BG temperature} to be $k_B T_{BG} \equiv 2\hbar v_p k_F$.) Above $T^*_{BG}$, in the so-called ``equipartition" (EP) regime, we instead expect linear-in-T resistivity. Single-layer graphene is a textbook example, displaying these features elegantly with $\mathcal{C}_{BG} \approx 1/6$ \cite{SankarGrapheneKT1, Efetov_2010}. We note that in situations (e.g. normal metals) where the Debye temperature is much smaller than $T_{BG}$, the crossover temperature becomes the Debye temperature because it is the maximum allowed phonon energy \cite{AshcroftAndMermin}.

While phonon scattering in TBLG has been investigated previously using the Dirac cone approximation \cite{Wu2019_phonon}, the full Bistritzer-MacDonald band structure is much more elaborate, containing Van Hove singularities, Lifshitz transitions, multibands, and a wide range of particle velocities and Fermi surface geometries. All of these features considerably complicate phonon-limited transport as detailed in Fig.~\ref{IntroductionDoSFigure}, and cannot be captured qualitatively or quantitatively by the Dirac cone approximation.
In this work, we present a theoretical treatment of acoustic phonon scattering induced electrical resistivity in TBLG, focusing on the effects of the detailed geometric features of the Bistritzer-MacDonald band structure.  

The purpose of this paper is to give a concrete calculation of the transport properties resulting from acoustic phonon scattering in TBLG. We give predictions for the doping ($n$) and temperature ($T$) dependence of the resistivity of these systems in the limit of phonon-dominated transport. Tuning the twist angle can significantly alter the band structure, including not only the Fermi velocity and the bandwidth, but also the location of the Van Hove singularities, affecting both the SC and the many correlated insulating phases. We therefore also predict the twist angle dependence of the transport properties arising from the acoustic phonon scattering.

Our calculations are done in the framework of Boltzmann kinetic theory, and we model the acoustic phonons via the Debye approximation. We retain the full electronic band structure obtained by the diagonalization of the Bistritzer-MacDonald Hamiltonian. We numerically solve the linearized Boltzmann equation in the anisotropic band geometry and quantitatively calculate the resistance. In particular, this allows us to identify various scaling regimes for the resistivity and the BG crossover temperature, $T^*_{BG}$ - see Fig.~\ref{TScalingPhaseDiagramFigure}. Treating the non-isotropic band structure in TBLG correctly leads to significant technical complication, beyond the techniques of prominent earlier analytical treatments of resistivity in 2D layered graphene structures \cite{SankarGrapheneKT1, SankarGrapheneKT2, Wu2019_phonon, Li2020, SankarGrapheneKT5}. 

The complicated electronic structure of the BM Hamiltonian for TBLG causes significant departures from the usual BG picture. While the EP regime behavior of the scattering rate of an individual Bloch state is linear, $1/\tau_{\vex{k}} \propto k_B T$, band curvature effects cause a complicated non-linear T-dependence of the resistivity. This is demonstrated in Figs.~\ref{TScalingPhaseDiagramFigure}-\ref{SecondarySlicesFigure}. Further, we note that the anisotropy in the band structure alters the low-$T$ BG relaxation rate $T^4$ power law to a non-universal, $\vex{k}$-dependent $T$-dependence. While this nonlinear-in-$T$ equipartition-regime phonon-limited resistivity defies the norm in the context of Boltzmann theory, we note that it has been detected experimentally in TBLG systems \cite{Polshyn2019}.

The debate over the presence of a strange metal phase in TBLG is but one important area of relevance for our current results. More generally, the recent progress in synthesizing 2D layered van der Walls heterostructures has brought this novel class of materials to the frontier of condensed matter physics\cite{Geim_2013, Novoselov_2006, Bistritzer_2011, Morell_2010, Li_2019, Kim_2017, Cao_2018a, Cao_2018b, Cao2020PRL, Cao2021, Yankowitz_2019,Kerelsky_2019, Lu_2019, Stepanov2020untying, Sharpe_2019, Chen_2020, Rozen2021entropic, AndreaYoungBernal, AndreaYoungRhombo, AndreaYoungRhombo2, Serlin_2020, Wu_2018, Wu_2019_TIPRL, KinFaiMak_TopologyTMD, Polshyn_2020, TBLGStrangeMetalExperiment1, Polshyn2019, TBLGStrangeMetalExperiment3, SankarFengchengStrangeMetal, CalTechBernal, CalTechSymmetryBreakingTBLG, Xie_2020, MacDonaldTLBGReview, Li_2021, Ghiotto_2021, Pan_2020, Pan_2021, Morales_Dur_n_2021, Ahn_2022, Kerelsky2021moireless, Khalaf2019, CalTechTwisted}. The sensitivity of the band structures of these materials to external control parameters (e.g. twist angles, external field) makes them an extremely versatile family of systems for realizing various exotic phases of 2D matter. Indeed, in addition to the possibly-exotic superconductivity \cite{Cao_2018a, Cao_2018b, Cao2020PRL, Cao2021, Yankowitz_2019,Kerelsky_2019, Lu_2019, Wu_2018, AndreaYoungRhombo2, AndreaYoungRhombo, AndreaYoungBernal, CalTechBernal, CalTechTwisted, CalTechSymmetryBreakingTBLG}, and possible ``strange metal" resistance scaling at very low temperature \cite{TBLGStrangeMetalExperiment1, Polshyn2019, TBLGStrangeMetalExperiment3, SankarFengchengStrangeMetal} reported in TBLG, 2D heterostructues have already shown various correlated insulating states \cite{Cao_2018a, Lu_2019}, ferromagnetism \cite{Sharpe_2019, Chen_2020}, correlation-driven valley and iso-spin polarization \cite{AndreaYoungBernal, AndreaYoungRhombo}, anomalous quantum Hall physics \cite{Serlin_2020}, topological insulator physics \cite{Wu_2019_TIPRL, KinFaiMak_TopologyTMD, Polshyn_2020}, metal-insulator transitions \cite{Li_2021, Ghiotto_2021, Pan_2020, Pan_2021, Morales_Dur_n_2021, Ahn_2022, Ahn_2022_Anderson}, and non-spin-singlet pairing superconductivity \cite{AndreaYoungBernal, AndreaYoungRhombo2}. In turn, this highlights the need for a refinement of the basic theories of phonon-limited resistivity as applied to these materials, accurately taking complex band geometry into account. Our results here for TBLG systems constitute a step in this direction. We emphasize that our work here centers on the electron-phonon interaction and ignores the effects of electron-electron interactions, which we do not believe play much of a role in determining the transport properties of TBLG in the metallic phase.

Our paper is organized as follows. We present an overview of the central results of our work in Sec.~\ref{Section-SummaryOfMainResults}, where we emphasize the most important quantitative aspects for comparison with experiment and qualitative results that run counter to common expectations. In particular, Sec.~\ref{Appendix-DiracConeComparison} discusses a quantitative comparison of our new results with the predictions of a Dirac cone approximation from previous work \cite{Wu2019_phonon}. We then provide a concise review of acoustic phonon scattering in kinetic theory and present an overview of the calculation of relaxation times in the TBLG system in Sec.~\ref{Section-ResistivityViaBoltzmannKineticTheory}. We emphasize the roles of anisotropy, band curvature, and moir\'{e} Umklapp scattering, all of which require more care than the standard isotropic, Umklapp-free case. For technical discussions of the role of the relaxation time approximation in solving the linearized Boltzmann equation and the iterative numerical solution of relaxation times, please see the appendices of Ref. \cite{Davis_2022}, where we developed the corresponding transport theory for untwisted, non-moire multilayer graphene, including full band structure effects. Our concluding discussion is presented in Sec.~\ref{Section-DiscussionAndConclusions}.




\section{Summary of main results}
\label{Section-SummaryOfMainResults}

Our central results are the calculations of the doping ($n$) and temperature ($T$) dependence of the resistivity [$\rho(n,T)$] for twisted bilayer graphene for several values of the twist angle ($\theta$). Our numerical results retain the full Bistritzer-MacDonald \cite{Bistritzer_2011} Hamiltonian and work under the assumption that scattering is limited to acoustic

\begin{figure*}[p!]
\includegraphics[angle=0,width=0.47\textwidth]{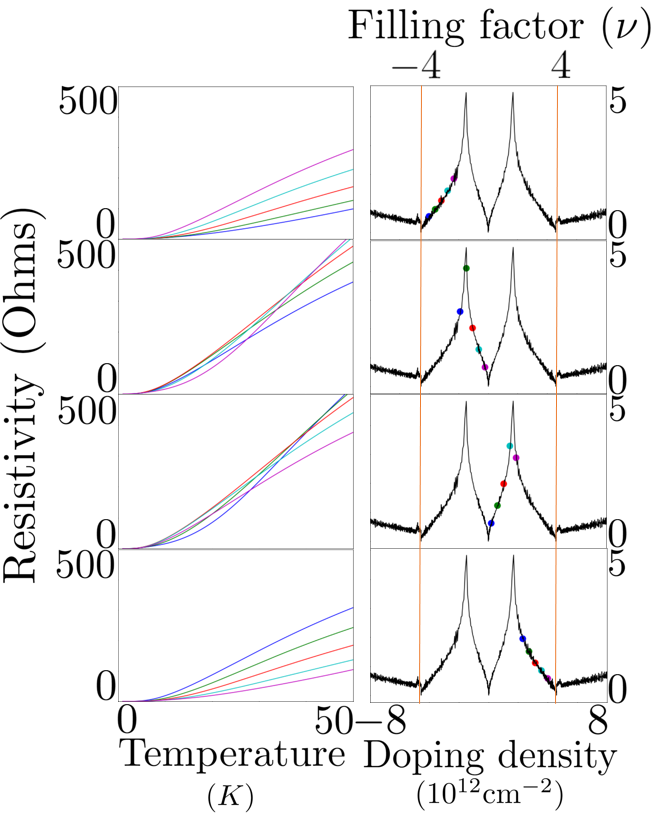}
\includegraphics[angle=0,width=0.48\textwidth]{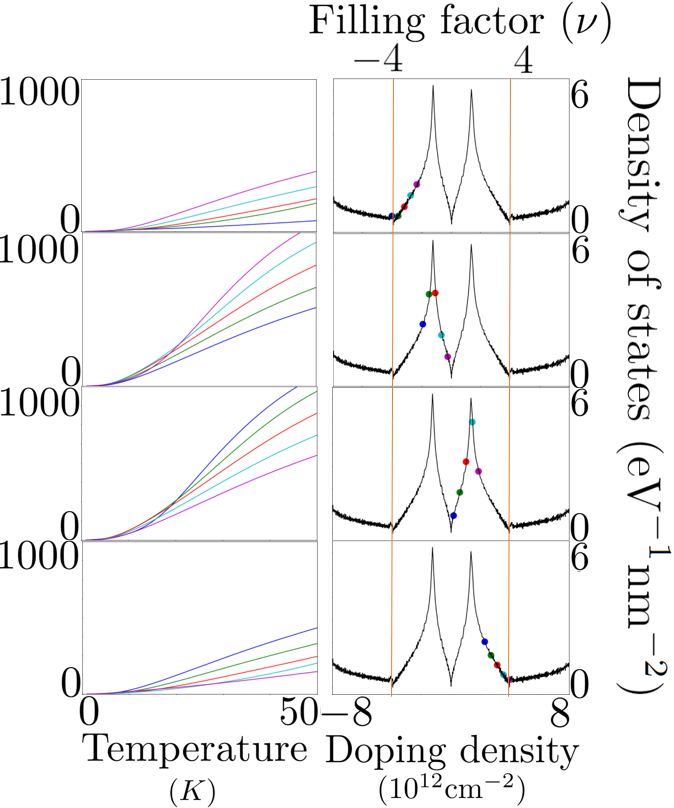}
\includegraphics[angle=0,width=0.48\textwidth]{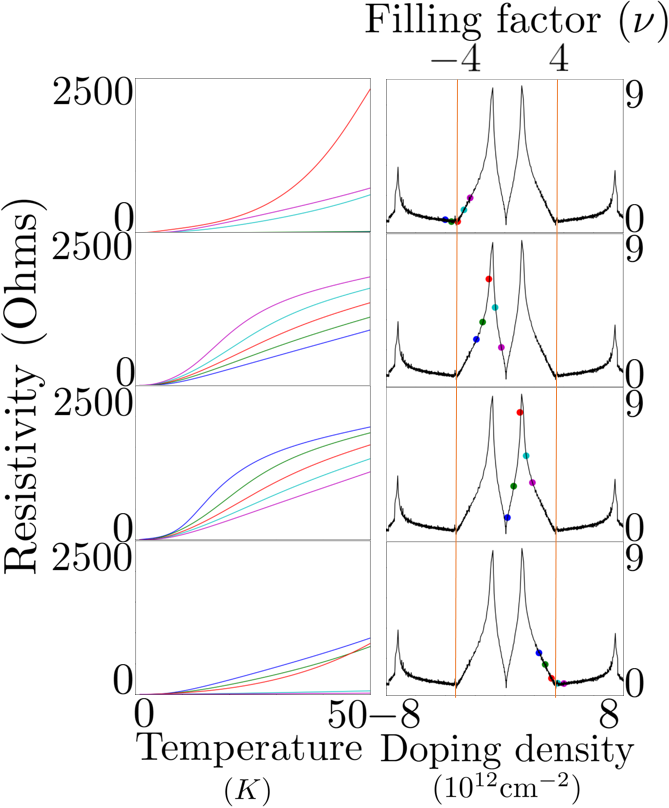}
\includegraphics[angle=0,width=0.49\textwidth]{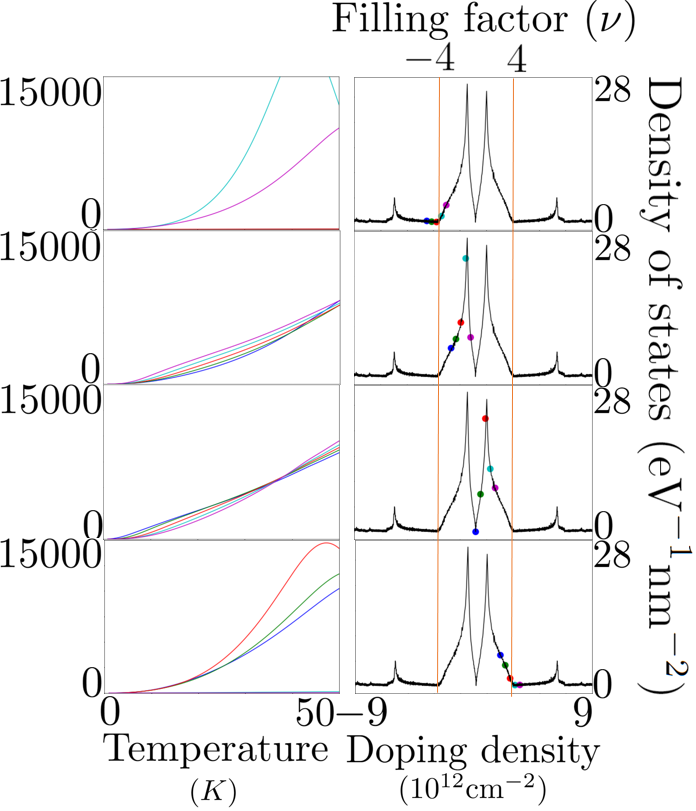}
\caption{Curves depicting the temperature ($T$) dependence of the resistivity [$\rho(n,T)$] for various dopings ($n$), focusing on the low-$T$ BG crossover regime. This figure recreates Figs.~\ref{PrimarySlicesFigure} and \ref{SecondarySlicesFigure}, but limits the scope to the low-$T$ regime. The upper left portion of the figure is for TBLG at twist angle $\theta = 1.4^{\circ}$, the upper right portion is for $\theta = 1.3^{\circ}$, the lower left portion is for $\theta = 1.2^{\circ}$, and the lower right portion is for $\theta = 1.1^{\circ}$. We plot resistivity over the temperature range [$0K,50K$], and we denote on the density of states plot which dopings the $\rho(T)$ curves correspond to.}
\label{BGCrossoverFigure}
\end{figure*}

\FloatBarrier

\begin{figure}[t!]
\includegraphics[angle=0,width=.47\textwidth]{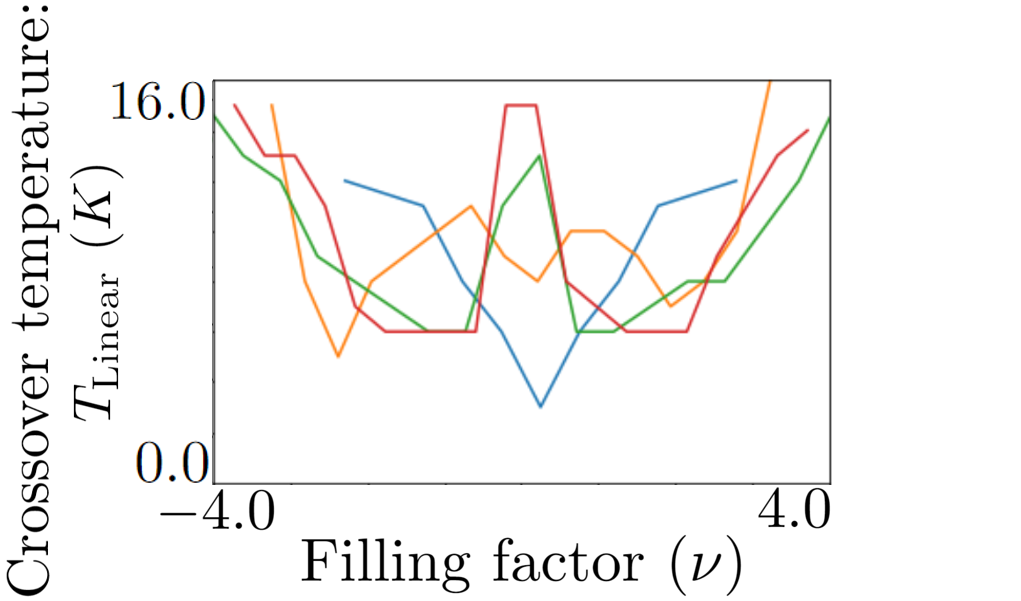}
\caption{We plot the effective crossover temperature to a regime of approximately linear resistivity scaling as a function of doping, for twist angles $\theta = 1.1^{\circ}$ (blue), $\theta = 1.2^{\circ}$ (orange), $\theta = 1.3^{\circ}$ (green), $\theta = 1.4^{\circ}$ (red). The crossover temperatures vary from around $5K$ to $16K$ for the various angles and dopings, with an exception for the data point very close to the Dirac point of the $\theta = 1.1^{\circ}$ model, where the crossover temperature is about $3K$. We emphasize that the BG crossover is generally not sharp, and in a complicated system like TBLG, the non-linearity in $\rho(T)$ due to band curvature competes with the BG process.}
\label{TLinearFigure}
\end{figure}

\begin{figure*}[t!]
\includegraphics[angle=0,width=\textwidth]{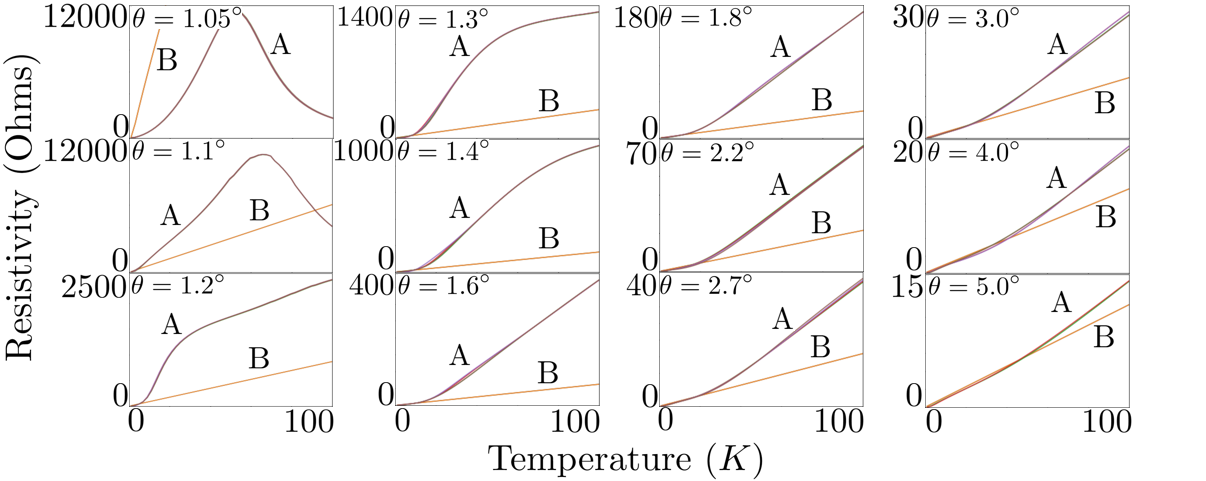}
\caption{Comparison of Dirac approximation to the full BM model. In each panel, we plot resistivity curves $\rho(T)$ for both (A) the BM model (for several doping values very near to the charge-neutrality point) and (B) the results of the corresponding Dirac cone approximation of Eq.~(\ref{SemiAnalyticDiracResistivity}) (for several doping values very near to the Dirac point). For large twist angles, we see that the Dirac cone approximation is accurate for temperatures up to $20K-50K$. For these systems, the departure from the Dirac cone approximation takes the form of a kink and sharp increase in slope, corresponding to the thermal activation of inter-cone scattering in the BM model. However, as the twist angle is decreased, the entire band structure becomes relevant for transport and the Dirac cone approximation is only valid for very low temperatures.}
\label{ComparisonWithWuDasSarmaFigure}
\end{figure*}

\noindent
phonons (which we treat in the Debye approximation.) In particular, we give quantitative predictions for the crossover from the Bloch-Gr\"{u}neisen regime to the equipartition regime. The key ingredient of the theory is the nonperturbative inclusion of the full TBLG band structure as obtained by BM in Ref. \cite{Bistritzer_2011}.

We plot $log[\rho(n,T)]$ for $T$ in ($0-300K$) for all twist angles under consideration in Fig.~\ref{LogRhoTFigure}. Individual curves of $\rho(n,T)$ for fixed $n$ are given in Figs.~\ref{PrimarySlicesFigure}. We note that the high-$T$ resistivity is not given by a simple $T$-linear power law above the BG regime. The resistivity is a complicated function of $n$ and $T$. We generally find resistivity peaks, followed by downturns in $\rho(n,T)$ which then either remain flat or return to linear dependence. While the nonlinear $T$-dependence and resistivity peaks are counter to high-$T$ phonon expectations based on simplistic theories, this behavior has already been reported in twisted bilayer \cite{Polshyn2019} graphene systems. This behavior is a manifestation of band curvature, and can be understood in systems as simple as a massive Dirac cone. In Sec.~\ref{Subsection-Peaks}, we show how to understand this physics in terms of kinetic theory, and this effect is discussed extensively in Ref.~\cite{Davis_2022}. We emphasize that these novel features of the theory are in general agreement with experimental observations.

We present data for $\theta = 1.4^{\circ}$ and $1.3^{\circ}$ in Fig.~\ref{PrimarySlicesFigure}. In both cases, we find gradual resistance peaks at $T \approx 100K$ as long as the filling factor $\nu$ falls in the range $-4 \le \nu \le 4$, where $\nu$ is the number of electrons per moir\'{e} unit cell. Explicitly, we have
\begin{align}
\label{FillingFactor}
    \nu \equiv n\mathcal{A}_0,
\end{align}
where $n$ is the doping density (units of $\text{length}^{-2}$) and $\mathcal{A}_0$ is the area of the moir\'{e} unit cell. Since we have a four-fold degeneracy due to spin and valley, the edges of the first moir\'{e} conduction and valence band are at $\nu = \pm 4$. (We show results for $\nu$ varying between up to $\pm 4$ in Fig.~\ref{PrimarySlicesFigure}). The height of the peak is relatively insensitive to doping. When the Fermi level is doped outside the first conduction (valance) moir\'{e} band range ($|\nu| > 4$) we do not see pronounced resistivity peaks. Such smooth resistivity maxima in temperature in the resistivity are generically observed in all TBLG transport experiments.

Similar results are given for $\theta = 1.2^{\circ}$ and $1.1^{\circ}$ in Fig.~\ref{SecondarySlicesFigure}, where we see the same qualitative features as in Fig.~\ref{PrimarySlicesFigure}. Doping levels in the range $-4 \le \nu \le 4$ give resistivity curves with low-$T$ peaks, with sharper peaks than found for the larger twist angle examples in Fig.~\ref{PrimarySlicesFigure}. The systems corresponding to lower twist angles (flatter bands) give sharper peaks. These examples also show that the magnitude of the resistivity is mostly unaffected by the doping within $|\nu| \le 4$; the exception is doping levels very close to the band edge, $|\nu| \approx 4$, which have a sharper and more pronounced peak. We point out that the sharp increase in resistivity for dopings very close to the band edge has also been seen in experiments, \cite{Polshyn2019}. Finally, we emphasize that as the bands flatten, the overall magnitude of the resistivity increases dramatically, varying over an order of magnitude over the twist angles that we study. This provides a possible clue for why earlier, simpler transport theories using the Dirac cone approximation necessitated an arbitrary upward adjustment of the electron-phonon coupling constant in order to obtain a quantitative agreement with the experimental TBLG transport data \cite{Wu2019_phonon}.

In Fig.~\ref{TScalingPhaseDiagramFigure}, we plot $d\log[\rho(n,T)]/d\log(T)$ as an approximate scaling exponent for the temperature-dependent power-law of the resistivity. These plots act as an effective ``phase diagram" for the various regimes of $T$-dependence in the resistivity profile. Corroborating our discussion above, we find there is a region where the resistivity curve flattens out to be essentially constant with $T$, sometimes after a downturn. Further, these plots show the Bloch-Gr\"{u}neisen crossover, where the $T$-scaling of the resistivity approaches $T^4$ at low $T$. Figure~\ref{TScalingPhaseDiagramFigure} shows that for all twist angles, the BG region (red) sets in for $T$ around $10K$, and this is essentially doping-independent throughout the flat band. However, it is also clear that the sharpness of the crossover regime can change dramatically with doping, as the geometry of the Fermi surface changes. In particular, as we dope the sample near the edge of the flat band, the sharp BG crossover evolves to one that interpolates between the BG and EP behavior over a temperature range spanning $100K$. In general, however, we do not find a situation where the linear-in-$T$ TBLG resistivity persists to a temperature much less than $10K$, e.g., never to below $1K$ for any doping or twist angle.

\subsection{Additional Results on BG crossover}
\label{Appendix-AdditionalData}
In Fig.~\ref{BGCrossoverFigure} we plot our calculated resistivity for the low-temperature regime containing the BG crossover. This figure mirrors Figs.~\ref{PrimarySlicesFigure} and \ref{SecondarySlicesFigure}. These results clarify our predictions of the typical temperatures at which a BG crossover should be expected. We see an essentially doping-independent BG crossover that takes place roughly near to $10K$ for all twist angles and doping levels under consideration. We explicitly give estimates of the crossover temperature to linear resistivity in Fig.~\ref{TLinearFigure}. These results make it clear that for small twist angles $\theta \in (1.1^{\circ},1.4^{\circ})$, and generic filling factor $|\nu| \leq 4$, we see a crossover to $T^{a>1}$ decay below a BG crossover temperature that ranges from $5-15 K$. Only for dopings extremely close to the Dirac point do we see a suppression of the BG temperature below $5K$, and this is in line with the standard theory of transport in Dirac systems \cite{Wu2019_phonon, SankarGrapheneKT1, SankarGrapheneKT2}. We emphasize that it is difficult to assign a precise value to the crossover temperature to the linear-in-$T$ resistivity regime in TBLG. In general, even in the simplest systems the $BG$ crossover to the equipartition regime is never sharp. Further, in the case of TBLG, we see that band-curvature effects lead to a highly nonlinear dependence of $\rho(T)$ on $T$, even in the equipartition regime. In some cases, these crossovers take place at the same temperature and it is impossible to delineate them from each other from resistivity data. Figure~\ref{TLinearFigure} represents the best estimate at a crossover temperature possible from our data. The important point is that the crossover regime is generally contained in the interval between $5$ and $15K$ for the twist angles we study.

\subsection{Comparison with Dirac cone approximation}
\label{Appendix-DiracConeComparison}

Earlier results on phonon-induced resistivity in TBLG have predicted giant linear-in-T resistance using only the Dirac cone approximation \cite{Wu2019_phonon}. The new theoretical idea underlying this resistivity enhancement in TBLG, which agrees with experiments, is the strong suppression of the effective Dirac velocity in TBLG due to the band flattening in the moir\'{e} system. This earlier work \cite{Wu2019_phonon} uses the BM model for TBLG to extract $v^*_F$ near the Dirac point for various twist angles. Once $v^*_F$ is extracted, Ref. 10 followed the theory of Ref. 17 (for monolayer graphene, but with the reduced Fermi velocity, $v^*_F$) to get a simple analytical result for the scaling of the resistivity in a Dirac system:

\begin{align}
\label{SemiAnalyticDiracResistivity}
    \rho(\theta,T,n) = \frac{F(\theta)}{v^*_F(\theta)^2}\frac{4 D^2 k_F}{e^2\rho_Mv_{p}}I\left(\frac{T}{T_{BG}}\right),\\
    I(z) = \frac{1}{z}\int_0^1 dx\ x^4\sqrt{1-x^2} \frac{e^{x/z}}{(e^{x/z}-1)^2}.
\end{align}
Above, $v_F^*(\theta)$ is the twist-angle-dependent flatband Fermi velocity and $F(\theta)$ is a form factor accounting for changes in the electron-phonon matrix element. (See \cite{Wu2019_phonon} for details.) The other parameters above are explained in Sec.~\ref{Subsection-Model}.

In this subsection, we compare our results to this earlier Dirac cone theoretic calculation \cite{Wu2019_phonon} - our current results use the same model parameters for the electron and phonon bandstructures and couplings as in Ref.~\cite{Wu2019_phonon}. We numerically extract $v_F^*$ from the BM model and compare Eq.~(\ref{SemiAnalyticDiracResistivity}) for the Dirac model with the temperature dependence of the resistivity of the full BM system, doped very close to the Dirac point. These results are given in Fig.~\ref{ComparisonWithWuDasSarmaFigure}. Here we also compare our small-angle results with larger twist angles, showing that the Dirac cone approximation holds to much higher temperatures for larger twist angles. This is expected since the moir\'{e} band structure effects are dominant only for lower twist angles, and the Dirac cone approximation improves with increasing twist angle. These results offer an understanding of the limitations of the Dirac cone approximation in fully capturing the transport physics at low temperatures as the twist angle is lowered. We emphasize that at very low temperatures, the two theories coincide. For angles as large as $5.0^{\circ}$, the Dirac cone model is valid up to around $50\ K$. However, for small twist angles in $1.1^{\circ}-1.4^{\circ}$, the theories start to differ at as low as a few Kelvin. We also note that for large twist angles, the departure from the Dirac cone model takes the form of a sharp kink in the slope of the resistivity curve, followed by another regime of linear scaling, corresponding to the thermal activation of inter-cone scattering in the BM band structure. This upward kink explains why the observed low-angle TBLG resistivity is much higher than the Dirac cone approximation results (necessitating an increase in the effective electron-phonon coupling parameter in Ref. \cite{Wu2019_phonon}). At twist angles closer to the magic angle condition, the Dirac cone approximation becomes inaccurate, and the whole band structure is important for the transport calculation even at low temperature. 

Our transport theory based on the full BM band structure is able to offer several significant improvements over the simplified transport theory of the Dirac cone model. Similar to Dirac cone model, the full BM theory predicts a crossover at low temperatures to a roughly linear-in-$T$ regime as the relaxation times enter the equipartition regime (see Sec.~\ref{Subsection-BGCrossover}), as observed experimentally \cite{Polshyn2019}. However, beyond the scope of the Dirac model, the band curvature of the BM model causes nonlinear behavior at higher temperatures. In particular, this gives the resistance peaks located at around $100K$, with the resistivity starting to decrease at higher T, as seen in experimental measurements \cite{Polshyn2019} (which has no explanation whatsoever within the Dirac cone model, where the resistivity would continue increasing as linear-in-$T$ forever). Figure~\ref{ComparisonWithWuDasSarmaFigure} shows that even very close to the Dirac point, the BM band structure still has a nontrivial effect and leads to significant enhancement of the resistivity over the Dirac cone result. We emphasize that this enhancement of the resistivity brings the theoretical results into much better agreement with experiments without any need for arbitrary adjustments of the deformation potential coupling. In particular, the resistivity peak at approximately $10k\Omega$ and $70K$ for the $1.1^{\circ}$ system is in excellent agreement with experimental results for small twist-angle systems \cite{Polshyn2019}. Finally, our model is also able to predict the observed enhancement of the resistivity at the filling factors $|\nu| \approx 4$, near the edge of the first moir\'{e} conduction and valence bands. Thus, the theory accounts both for the large enhancement in the resistivity as well as its slow decrease at higher temperatures.




\section{Resistivity via Boltzmann kinetic theory}
\label{Section-ResistivityViaBoltzmannKineticTheory}
Boltzmann kinetic theory (BKT) is a powerful and well-established theoretical technique for the calculation of linear response resistivities \cite{AshcroftAndMermin, Ziman, SankarGrapheneKT1, SankarGrapheneKT2}. In this section, we outline the application of BKT to the problem of collisions between acoustic phonons and Bloch state electrons in twisted bilayer graphene. We introduce the model in Sec.~\ref{Subsection-Model} and state the central formulae of the kinetic theory in Sec.~\ref{Subsection-BasicTheory}. In Sec.~\ref{Subsection-BGCrossover}, we provide the physical intuition for the Bloch-Gr\"{u}neisen (BG) and equipartition (EP) scattering regimes and then in Sec.~\ref{Subsection-Anisotropy} we explain how these paradigms are altered to account for band anisotropy. In Sec.~\ref{Subsection-Umklapp}, we discuss the enhanced role of Umklapp scattering due to the small size of the moir\'{e} Brillouin zone. Finally, we give an overview of the resistivity computation protocol in Sec.~\ref{Subsection-NumericalComputation}.

\subsection{Model}
\label{Subsection-Model}
Our electron-phonon model is described by the Hamiltonian
\begin{align}
\label{TotalHamiltonian}
    H = H^{e} + H^{ph} + H^{e-ph},
\end{align}
where the single-particle electron part is given by the Bistritzer-MacDonald (BM) Hamiltonian \cite{Bistritzer_2011}
\begin{align}
\label{BMHamiltonian}
    H^e &\equiv 
    \sum_{\vex{k},\vex{k'}}
    c^\dagger_{\vex{k'}}
    H^{BM}_{\vex{k'},\vex{k}}
    c_{\vex{k}},
    \\
    &=
\label{BMDiagonal}    
    \sum_{b,\vex{k}\in\text{MBZ}}
    \e^{BM}_{b,\vex{k}}
    \tilde{c}^\dagger_{b,\vex{k}}
    \tilde{c}_{b,\vex{k}},
\end{align}
where $\mathcal{A}$ is the system area and $c_{\vex{k}}^\dagger \equiv c^\dagger_{s,\xi,\sigma,l,\vex{k}}$ creates an electron with momentum $\vex{k}$ (relative to the Dirac point in the continuum model of graphene), spin $s$, valley $\xi$, sublattice $\sigma$, and layer $l$.
\begin{align}
    H^{BM}_{\vex{k'},\vex{k}} \equiv \delta_{s,s'}\delta_{\xi,\xi'}H^{BM}_{\sigma',l',\sigma,l,\vex{k'},\vex{k}}
\end{align} 
is a matrix coupling together layer, sublattice, and momentum degrees of freedom through a periodic moir\'{e} potential. Our convention is that sums over unwritten indices $\{s,\xi,\sigma,l\}$ are implicit. In Eq.~(\ref{BMHamiltonian}), the $\vex{k}$ summation is unbounded, reflecting the continuum limit in the Bistritzer-MacDonald treatment of the constituent graphene layers. (However, a high-energy cutoff is re-introduced in our numerical calculations.) In Eq.~(\ref{BMDiagonal}), we introduce the moir\'{e}-Bloch functions
\begin{align}
\label{MoireBloch}
\tilde{c}^\dagger_{s, \xi, b,\vex{k}} = \sum_{\sigma, l, \vex{G}} V_{b,\vex{k}; \sigma, l, \vex{G}}
\ c^\dagger_{s,\xi,\sigma,l,\vex{k}+\vex{G}},
\end{align}
where $\vex{k}$ is in the moir\'{e} Brillouin zone, $\vex{G}$ runs over the reciprocal lattice, and $b$ is the band index of the BM eigenfunction. The basis-change matrix $V_{b,\vex{k}; \sigma, l, \vex{G}}$ defines the representation of the Bloch wavefunctions in the moir\'{e} Brillouin zone crystal momentum basis. The four degenerate spin-valley flavors remain decoupled in our calculation and contribute equally to the conductivity (inverse resistivity). We will generally stop referencing them in the following. We use the standard BM Hamiltonian with the interlayer hopping parameters $\omega_0 = 90\ meV$, $\omega_1 = 117\ meV$ and the bare graphene Dirac cone velocity $v_F = 10^6 m/s$, which places the ``magic angle" at $\theta = 1.025^{\circ}$ \cite{Wu2019_phonon}.

We are primarily interested in the effects of the geometry of the BM band structure, so we restrict our model to in-plane longitudinal acoustic phonons and adopt a simple Debye description. We thus take the phonon Hamiltonian to be
\begin{align}
\label{PhononHamiltonian}
H^{ph} = \sum_{l,\vex{q}}\hbar\omega_{\vex{q}}a_{l,\vex{q}}^\dagger a_{l,\vex{q}},
\end{align}
where $\omega_{\vex{q}}$ is the phonon dispersion and we use the Debye approximation $\omega_{\vex{q}} \approx v_{p}|\vex{q}|$, where $v_p$ is the phonon (or sound) velocity of graphene. In turn, the phonons couple to the electrons via the deformation potential coupling Hamiltonian \cite{Ziman, SankarGrapheneKT1, Coleman2015introduction}:
\begin{align}
\label{CouplingHamiltonian}
    H^{e-ph} &= 
    \sqrt{\frac{D^2 \hbar}{2\rho_M \mathcal{A}}}
    \sum_{l,\vex{q}}
    \frac{\hat{n}_{\vex{q},l}}{\sqrt{\omega_{\vex{q}}}}(-i\vex{q}\cdot\hat{e}_{\vex{q}})(a_{\vex{q},l} + a_{-\vex{q},l}^\dagger).
\end{align}
Here, $D$ is the deformation potential, $\rho_M$ is the mass density of monolayer graphene, and $\hat{e}_{\vex{q}}$ is the desplacement unit vector of the phonon. Throughout this work, we set $D = 25$ eV, $\rho_M = 7.6 \cdot 10^{-8} g/cm^2$, and $v_p = 2.0 \cdot 10^6 cm/s$, following the standard graphene literature \cite{SankarGrapheneKT1, SankarGrapheneKT2, Wu2019_phonon, Efetov_2010}. Finally, the electron density operator is 
\begin{align}
\label{DensityOperator}
    \hat{n}_{\vex{q},l} \equiv \sum_{\vex{k}}c^{\dagger}_{(\vex{k}+\vex{q}),l}c_{\vex{k},l}.
\end{align}
As before, sums over $s,\xi,$ and $\sigma$ are implicit in Eq.~(\ref{DensityOperator}).

\subsection{Kinetic theory}
\label{Subsection-BasicTheory}

The ``relaxation time approximation" \cite{AshcroftAndMermin} to Boltzmann kinetic theory (BKT) gives a simple formula for the resistivity tensor ($\rho$):
\begin{align}
\label{ResistivityDefinition}
    [\rho^{ij}(n,T)]^{-1} 
    &= 
    \frac{4e^2}{\mathcal{A}}
    \sum_{S}
    \tau_{S}
    v^i_{S}v^j_{S}
    \partial_{\mu}f(\e_S),
\end{align}
where $T$ is temperature, $e$ is the electron charge, $S \equiv \{b,\vex{k}\}$ denotes a moir\'{e}-Bloch state, $v_{S}^j$ are components of the velocity of the state $S$, $\e$ is the energy of $S$, $f(\e)$ is the Fermi distribution function, and the $\tau_{S}$ are the state-dependent \textit{relaxation times} of the various Bloch states. The summation on $S$ in Eq.~(\ref{ResistivityDefinition}) runs over all moir\'{e}-Bloch states. If the band structure and moir\'{e}-Bloch states are known, the main challenge in the calculation of the resistivity is the computation of the relaxation times. The factor of $4$ follows from the spin and valley degeneracies of the TBLG.

A standard ``Fermi's golden rule" calculation - assuming a thermal equilibrium distribution for the phonons - gives the scattering rate between Bloch states $S$ and $S'$ as 
\begin{align}
    \nonumber
    \mathcal{W}_{S \rightarrow S'}
    &=
    \frac{\pi D^2 |\vex{q}|}{\rho_M v_p}
    \Delta_{S,S'}
    \sum_{l}
    \bigg|\langle S'|\hat{n}_{l,\vex{q}}|S\rangle\bigg|^2,
\\
\label{TransitionRates}
    &\equiv
    \hbar v_p|\vex{q}|
    \Delta_{S,S'}
    \mathcal{C}_{S,S'},
\end{align}
with 
\begin{align}
    \vex{q} &\equiv \vex{k'}-\vex{k}+\vex{Q},
    \\
\label{EnergyConservation}    
    \Delta_{S,S'}
    &\equiv
    \begin{aligned}
    N_{\vex{q}} &\delta(\e'-\e - \hbar v_p |\vex{q}|) 
    \\
    + 
    (N_{\vex{q}} + 1 ) &\delta(\e'-\e + \hbar v_p |\vex{q}|)
    \end{aligned}
    ,
    \\
\label{OccupationNumber}
    N_{\vex{q}} &\equiv \frac{1}{\exp(\hbar v_p |\vex{q}|/k_BT) - 1},
\end{align}
for some moir\'{e} reciprocal lattice vector $\vex{Q}$. The Dirac $\delta$-functions in Eq.~(\ref{EnergyConservation}) enforce conservation of energy and moir\'{e} lattice momentum and $N_{\vex{q}}$ gives the occupation numbers of phonons available for scattering. The first line in Eq.~(\ref{EnergyConservation}) refers to phonon absorption processes while the second refers to phonon emission. 

\begin{figure*}[t!]
    \centering
    \begin{minipage}{0.50\textwidth}
        \centering
        \includegraphics[width=\textwidth]{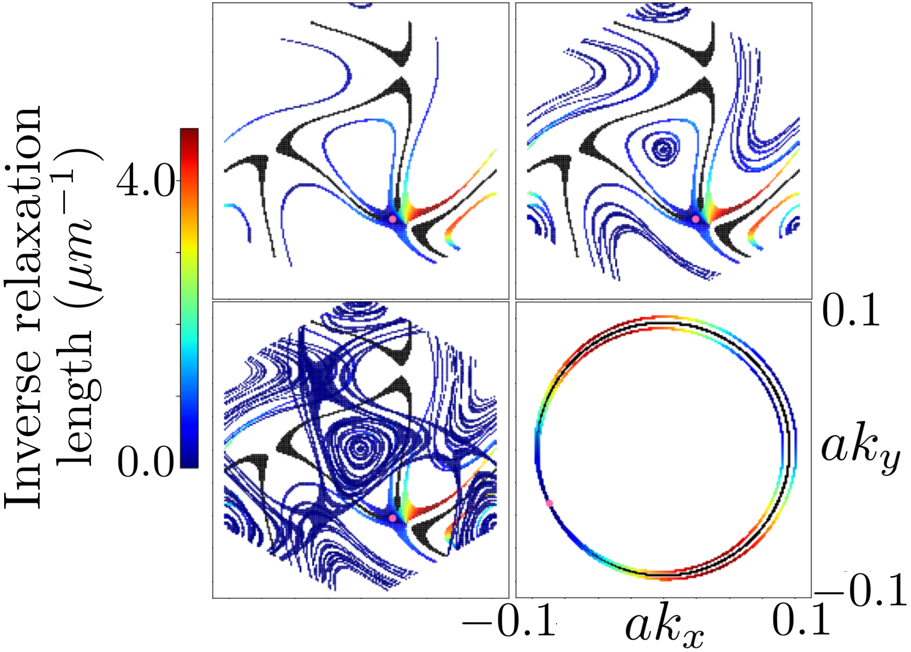}
    \end{minipage}
    \begin{minipage}{0.47\textwidth}
        \centering
        \includegraphics[width=\textwidth]{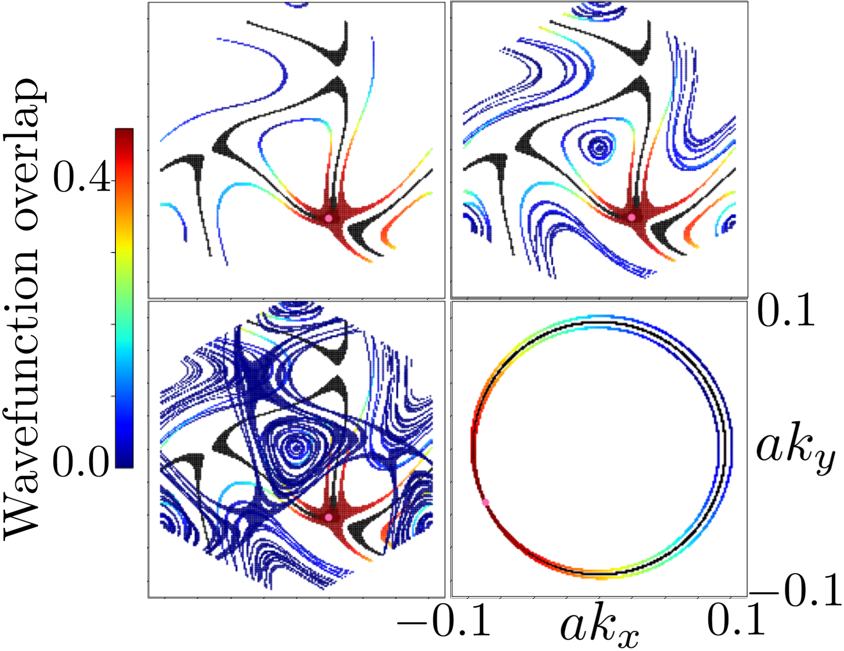}
    \end{minipage}
    \caption{Plots of kinematically allowed scattering manifolds for TBLG and a simple Dirac cone (for comparison). These are the points picked out by the energy-momentum conserving delta functions in Eq.~(\ref{SMCondition}). In each image above, a reference state is marked in pink, the corresponding Fermi surface is plotted in black, and the kinematically allowed scattering manifold (SM) is plotted in color. In each figure, the top left, top right, and bottom left panels depict scattering manifolds for $\theta = 1.3^{\circ}$ twisted bilayer graphene, with a reference state at $\mu = 0.012\ eV$. The top left panel neglects Umklapp scattering entirely, the top right panel allows Umklapp scattering only to the ring of adjacent Brillouin zones, and the bottom left panel allows Umklapp scattering to the first two rings of adjacent Brillouin zones. The bottom right panel shows the SM of a simple Dirac cone, at $\mu = -0.2\ eV$. All panels are plotted over the region $a(k_x,k_y) \in [-0.1,0.1]^2.$ In the figure on the left, the SM is color coded to indicate the scattering rate from the reference state to each state on the SM (The scattering rates have been computed at $100\ K$). On the right, the SM is color coded to indicate the wavefunction overlap between the reference state and each state on the SM. In the Dirac cone example, we can see that conservation laws provide a hard cutoff to the states available for scattering. Conversely, in the TBLG example, we see that small-momentum moir\'{e}-Umklapp scattering provides access to kinematically allowed scattering states for a wide range of energies, and there is not a hard cutoff imposed by kinematic constraints. In the TBLG, the decay of the wavefunction overlap term is thus necessary for a Bloch-Gr\'{u}neisen crossover.
}
    \label{ScatteringManifoldFigure}
\end{figure*}

The conservation laws in Eq.~(\ref{EnergyConservation}) determine a \textit{scattering manifold} for each Bloch state, defining the set of final states that can be scattered to without violation of energy or moir\'{e} crystal momentum. Examples of scattering manifolds are depicted in Fig.~\ref{ScatteringManifoldFigure}, where the TBLG case is compared with the much simpler case of a Dirac cone.

Noting that 
\begin{align}
    \hbar v_p |\vex{q}| \Delta_{S,S'} = \frac{\e'-\e}{\exp[(\e'-\e)/k_B T]-1} \tilde{\Delta}_{S,S'},
\end{align}
with
\begin{align}
\label{SMCondition}
     \tilde{\Delta}_{S,S'} \equiv \delta(|\e'-\e|-\hbar v_p |\vex{q}|),
\end{align}
and enforcing self-consistency of the relaxation time approximation on the Boltzmann equation, we find that
\begin{align}
    \label{RelaxationLengthSelfConsistency}
    \frac{1}{|\vex{v}_{S}| \mathcal{A}}
    \sum_{S'}
    \tilde{\Delta}_{S,S'}
    \mathcal{C}_{S,S'}
    \mathcal{F}_{S,S'}^{\mu,T} 
    \left[
    l_{S} - l_{S'}\cos\theta_{\vex{v}}
    \right]
    &= 1,
\end{align}
where $l_{S} \equiv |\vex{v}_{S}|\tau_{S}$ are the ``relaxation lengths" (mean free paths), $\theta_{\vex{v}}$ is the angle between the Bloch velocities $\vex{v}_{S}$ and $\vex{v}_{S'}$. As in Eq.~(\ref{ResistivityDefinition}), the summation over $S'$ is over all moir\'{e}-Bloch states and $\tilde{\Delta}_{S,S'}$ restricts the summation to the scattering manifold. We also have defined the function
\begin{align}
    \mathcal{F}_{S,S'}^{\mu,T}
    \equiv 
    \frac{1-f(\e')}{1-f(\e)}
    \frac{(\e'-\e)}{\exp[(\e'-\e)/k_BT])-1}
\end{align}
which encodes all implicit dependence of the relaxation lengths on the temperature or chemical potential.
 
In the thermodynamic limit, Eq.~(\ref{RelaxationLengthSelfConsistency}) becomes an integral equation. For a finite-size system, it is a matrix equation that can be inverted to find the relaxation lengths \cite{Davis_2022}. Solving this integral equation is the fundamental problem in the BKT approach to transport.

\subsection{Bloch-Gr\"{u}neisen regime}
\label{Subsection-BGCrossover}

Here we provide some intuition for the Bloch-Gr\"{u}neisen power-law regime at low temperature in the standard case. We consider a system that is isotropic ($l_{\vex{k}} \rightarrow l_{\e_{\vex{k}}}$ and $\vex{v}_{\vex{k}} \parallel \vex{k}$) and quasi-elastic ($\e' \approx \e$), and we assume there is only a single band. In this case, we can replace the velocity angle with the momentum angle ($\theta_{\vex{v}} = \theta_{\vex{k}}$) and can refer to a state $S$ by its lattice momentum $\vex{k}$. With these assumptions, Eq.~(\ref{RelaxationLengthSelfConsistency}) reduces to a direct formula for the relaxation time: 
\begin{align}
    \label{IsotropicRelaxationTime}
    \frac{1}{\tau_{\vex{k}}}
    &=
    \frac{1}{\mathcal{A}}
    \sum_{\vex{k'}}
    \tilde{\Delta}_{\vex{k},\vex{k'}}
    \mathcal{C}_{\vex{k},\vex{k'}}
    \mathcal{F}_{\vex{k},\vex{k'}}^{\mu,T}
    \left[
    1 - \cos\theta_{\vex{k}}
    \right],
\end{align}
where $\tilde{\Delta}_{S,S'} \approx 2\delta(\e'-\e)$ essentially restricts the summation to the Fermi surface, which is taken to be indistinguishable from the scattering manifold in the quasi-elastic approximation.

Now we consider the limit of very low temperatures. For low $T$, the phonon occupation function $N_{\vex{q}}$ (via $\mathcal{F}^{\mu,T}_{S,S'}$) strongly suppress $\vex{k'}$ in the sum in Eq.~(\ref{IsotropicRelaxationTime}) that involves large momentum transfer. We are left with $\vex{k'}$ such that $\hbar v_p |\vex{q}| \leq k_B T$. Therefore, in the summand of Eq.~(\ref{IsotropicRelaxationTime}), we may expand in small $\vex{q}$. Doing this, we see that $1-\cos\theta_{\vex{k}} \approx |\vex{q}|^2$ and $\mathcal{C}_{\vex{k},\vex{k'}} \approx 1$, and the summand of Eq.~(\ref{IsotropicRelaxationTime}) scales with $\vex{q}$ roughly as $|\vex{q}|^3$. 
Since the sum is effectively restricted to $\hbar v_p |\vex{q}| \leq k_B T$, the important contribution comes from the roughly spherical [$(d-1)$-dimensional] neighborhood of the scattering manifold with a radius proportional to $T$. Summing $|\vex{q}|^3$ over this sphere gives the famous power-law defining the BG regime:
\begin{align}
    \frac{1}{\tau_{\vex{k}}} \propto T^{d+2}.
\end{align}

\subsection{Equipartition regime}

The high-$T$ equipartition (EP) regime for the relaxation length $l_{S}$ sets in when
\begin{align}
\label{HighTExpansionCondition}
|\e'-\e| \ll k_BT
\end{align}
for all points $S'$ on the scattering manifold for the state $S$. In the isotropic case, this is simply the condition that 
\begin{align}
\label{IsotropicEPCondition}
    k_BT \gg k_B T_{BG} \equiv 2\hbar v_p k_F.
\end{align}
If $T_{BG} > T_{Debye}$, as is true for all normal metals, then $T_{Debye}$ replaces $T_{BG}$ in the inequality above. (This is not relevant for graphene where $T_{Debye} > 10^3 K$ and $T_{BG}$ is usually less than $50K$.) When Eq.~(\ref{HighTExpansionCondition}) is satisfied, we can expand in small $|\e'-\e|/(k_BT):$ 
\begin{align}
\label{FExpansion}
    \mathcal{F}_{S,S'}^{\mu,T} = k_B T + \mathcal{O}(\Delta\e/T).
\end{align}
Inserting this into Eq.~(\ref{RelaxationLengthSelfConsistency}) gives
\begin{align}
    \label{RelaxationLengthHighT}
    \frac{k_B T}{|\vex{v}_{S}| \mathcal{A}}
    \sum_{S'}
    \tilde{\Delta}_{S,S'}
    \mathcal{C}_{S,S'}
    \left[
    l_{S} - l_{S'}\cos\theta_{\vex{v}}
    \right]
    &= 
    1 + \mathcal{O}(\Delta\e/T)^2.
\end{align}
Solving Eq.~(\ref{RelaxationLengthHighT}) order-by-order in $1/T$, we see that the high-$T$ form of the relaxation length is 
\begin{align}
\label{EquipartitionLengths}
    l_{S} &= \frac{c_{S}}{k_B T} + \mathcal{O}(\Delta\e/T)^3,
\end{align}
where the $c_S$ give the asymptotic proportionality constant between the relaxation length and the inverse temperature in the high-$T$ limit. In the high-$T$ equipartition regime, all geometric information about the band structure relevant to the transport properties of the system is contained in the constant $c_S$.

We note that the $\mathcal{O}(1)$ term in the $\Delta\e/T$ expansion of $\mathcal{F}_{S,S'}^{\mu,T}$ in Eq.~(\ref{FExpansion}) rather remarkably vanishes, preventing a  $\mathcal{O}(\Delta\e/T)^2$ term in Eq.~(\ref{EquipartitionLengths}). This implies that the high-$T$ scattering rate (due to phonons) of a given Bloch state should be purely linear, going to zero in the $T \rightarrow 0$ extrapolation. 

We emphasize that the EP scaling law in Eq.~(\ref{EquipartitionLengths}) may set in at a physical crossover temperature $T^*_{BG},$ which could be lower than $T_{BG}.$  While $k_BT_{BG}$ defines the largest energy differences allowed in scattering by kinematic constraints, other terms in Eq.~(\ref{RelaxationLengthSelfConsistency}), such as the wavefunction overlap term $\mathcal{C}_{S,S'}$, can suppress large-energy scattering. This is demonstrated for the simple Dirac cone Hamiltonian in Fig.~\ref{ScatteringManifoldFigure}. Indeed, monolayer graphene displays a BG crossover to linear-in-T resistivity scaling at $T^*_{BG} \approx T_{BG}/6$ \cite{Efetov_2010, SankarGrapheneKT1, SankarGrapheneKT2}. This feature is more important in moir\'{e} systems, where multiple-Umklapp scattering events are possible due to the small moir\'{e}-Brillouin zone. Thus, $T_{BG}$ is simply a parametric crossover temperature scale above which the linear-in-$T$ equipartition applies, the real crossover is often at a temperature much lower than $T_{BG}$. We revisit this in Sec.~\ref{Subsection-Umklapp}.

\subsection{Alterations due to band anisotropy}
\label{Subsection-Anisotropy}

The BG crossover is more complicated with a non-isotropic system. First, in the low-$T$ limit, if we do \textit{not} assume isotropy, then we must restore
\begin{align}
\label{Anisotropy}
    1 - \cos\theta_{\vex{k}} \rightarrow 1 - \frac{l_{S'}}{l_{S}}\cos\theta_{\vex{v}}
\end{align}
in Eq.~(\ref{IsotropicRelaxationTime}). The small-$|\vex{q}|$ limit of the left-hand side of Eq.~(\ref{Anisotropy}) is simply proportional to $|\vex{q}|^2$. However, the small-$|\vex{q}|$ limit of the right hand side is more complicated since it also depends on the limits $l_{S'} \rightarrow l_{S}$ and $\vex{v}_{S'} \rightarrow \vex{v}_{S}$ as $S' \rightarrow S.$ We therefore expect anisotropy to introduce non-universal, state-dependent modifications of the BG power law in the $T$-dependence of each relaxation time $\tau_{S}.$ The isotropic Dirac cone approximation misses these subtleties.

Further, different points on a non-circular Fermi surface may have qualitatively different scattering manifolds, and therefore may cross into high-$T$ scaling at different thresholds. To discuss the high-$T$ limit, we must generalize the isotropic result in Eq.~(\ref{IsotropicEPCondition}).
We define the generalized (state-dependent) BG temperature as
\begin{align}
\label{TBGDefinitionGeneralized}
k_BT_{BG}(S) = \max_{S'\in\text{SM}(S)} |\e'-\e|.
\end{align}
The equipartition regime is the range of temperature for which Eq.~(\ref{EquipartitionLengths}) holds for all Block states $S$ in the thermally-active energy range around $\mu(n,T)$. Since the BG temperatures are state-dependent, we should generally expect a more gradual BG crossover than seen in isotropic systems. 

However, the linear-in-$T$ power law for the relaxation rate of the EP regime is not affected by anisotropy, unlike its BG regime counterpart. Eq.~(\ref{FExpansion}) does not depend on state-specific information, so as long as all states are in their BG scaling regime, Eq.~(\ref{EquipartitionLengths}) holds and all band structure information is encoded in the constants $c_{S}$, introduced in Eq.~(\ref{EquipartitionLengths}).

\subsection{Resistivity peaks and high-$T$ nonlinear $\rho(T)$}
\label{Subsection-Peaks}
We have seen in Eq.~(\ref{EquipartitionLengths}) that in the equipartition regime (asymptotic high-$T$ regime), the relaxation rate for each individual Bloch state scales linearly with temperature. This result is very general and holds for an arbitrary electronic band structure. It is commonly stated \cite{AshcroftAndMermin} that this implies that the high-$T$ resistivity is also linear in temperature, at least in the kinetic theory prediction. This is in fact not true due to the thermal averaging in Eq.~(\ref{ResistivityDefinition}).

More precisely, let us define the function
\begin{align}
\label{GFunctionDefinition}
    \delta^{ij}g(\e) \equiv \frac{1}{\mathcal{A}}\sum_{S} \frac{v^i_S v^j_S}{|\vex{v}_S|} c_S\ \delta[\e-\e_S],
\end{align}
where $c_S$ are the proportionality constants defined in Eq.~(\ref{EquipartitionLengths}). The equipartition regime resistivity is expressed simply in terms of $g(\e)$:
\begin{align}
\label{ResistivityFromG}
    \frac{1}{\rho(\mu,T)} = \frac{1}{(k_B T)^2}\int d\e\ g(\e) f(\e) [1-f(\e)].
\end{align}
From Eq.~(\ref{ResistivityFromG}), it is clear that we should expect linear-in-$T$ resistivity in the equipartition regime as long as the integral over $\e$ scales linearly with $T$. This is true, in particular, if $g(\e)$ is roughly linear as a function of energy in a neighborhood of width $k_B T$ about the chemical potential $\mu(n,T)$ (i.e $g''(\e)$ is small in the range $[\mu - k_BT, \mu + k_BT]$). For example, in the case of a Dirac cone band structure, one finds that $g(\e)$ is constant, and a linear-in-$T$ resistivity is robust at high $T$. On the other hand, if the integral in Eq.~(\ref{ResistivityFromG}) does not scale linearly with $T$, we can expect a more complicated dependence of the resistivity on $T$. This is possible if curvature in the band structure leads to non-linear behavior in $g(\e)$ In particular, $g(\e)$ necessarily vanishes in a band gap; we often find resistivity peaks or saturation when carriers near a band edge are thermally activated. This effect is explored extensively in Ref.~\cite{Davis_2022}. 

In TBLG, the narrow bandwidth of the first moir\'{e} conduction and valence bands cause the integral in Eq.~(\ref{ResistivityFromG}) to scale non-linearly with $T$ at relatively low $T$, explaining the observed resistance saturation. As expected, at larger twist angles, where the bandwidth is significantly larger, we observe linear-in-$T$ resistivity to higher temperatures (see Fig.~\ref{ComparisonWithWuDasSarmaFigure}). Previous work has conjectured that excitations of carriers in higher bands is responsible for the resistivity peaks in TBLG \cite{Polshyn2019}. We emphasize that while thermal excitations to higher bands can indeed contribute to nonlinear-in-$T$ resistivity in the equipartition, higher bands are not necessary for this physics. The nonlinearity and/or a resistivity peak arises naturally from the complex BM band structure without invoking higher bands.

\subsection{Moir\'{e}-Umklapp scattering in TBLG}
\label{Subsection-Umklapp}

In TBLG, the moir\'{e} Brillouin zone is much smaller than the Brillouin zone of regular (monolayer) graphene or a standard crystal lattice; near the ``magic angle", the moir\'{e} reciprocal lattice basis vectors (RLBV) have a length $a|\vex{Q}_{RLBV}|\approx 0.1$, where $a$ is the monolayer graphene lattice constant. This greatly enhances the importance of Umklapp scattering in the moir\'{e} zone. A phonon carrying the momentum of the RLBV will only have energy $\hbar v_p |\vex{Q}_{RLBV}| \approx 0.005 eV$ and these will be thermally active at low temperatures.

The availability of moir\'{e} Umklapp scattering causes TBLG to host huge scattering manifolds compared to those in non-moir\'{e} systems. Examples of scattering manifolds found in TBLG states are compared with those for simple Dirac cone graphene in Fig.~\ref{ScatteringManifoldFigure}. It is clear that the availability of essentially unlimited Umklapp scattering can make $T_{BG}$, as defined via Eq.~(\ref{TBGDefinitionGeneralized}) ill-defined. In this case, it is possible that the kinematical conservation laws do not provide the same sharp cut-off to the scattering manifolds of TBLG that they do in the standard picture. Instead, we must look to the decay of the overlap term,
\begin{align}
    \frac{\hbar \rho_M v_p^2}{\pi D^2}\mathcal{C}_{S,S'} = 
    \sum_{l}
    \bigg|\langle S'|\hat{n}_{l,\vex{q}}|S\rangle\bigg|^2,
\end{align}
to provide a smooth cutoff to the states ($S'$) that can meaningfully contribute to the scattering rate of $S$. We thus should understand the BG crossover in TBLG to be a more gradual process than that in non-moir\'{e} systems.

With regard to the inner product, we have
\begin{align}
    \langle S'|\hat{n}_{l,\vex{q}}|S\rangle = \sum_{\sigma,\vex{G},\vex{G'}}
    V^*_{b',\vex{k'}; \sigma, l, \vex{G}}
    V_{b,\vex{k}; \sigma, l, \vex{G'}}
    \delta_{\vex{k'}+\vex{G'},\vex{k}+\vex{G}+\vex{q}}.
\end{align}
We see that the $\delta$-function implements a sort of ``shift matrix" for the overlap of states connected by an Umklapp process. Figure \ref{ScatteringManifoldFigure} plots the magnitude of the overlap over a multi-Umklapp scattering manifold for TBLG, comparing it to the analogous situation in Dirac cone graphene. We note that even in Dirac cone graphene, the overlap term suppresses scattering to the far side of the Fermi surface and reduces the effective crossover temperature, $T^*_{BG}$. 

\subsection{Resistivity computation}
\label{Subsection-NumericalComputation}

Equations (\ref{ResistivityDefinition}-\ref{OccupationNumber}) combined with knowledge of the Bloch states give all the tools necessary to make a resistivity prediction. We solve Eqs.~(\ref{RelaxationLengthSelfConsistency}) for scattering lengths for each Bloch state. We emphasize that in general, the relaxation lengths $\{l_{\vex{k}}\}$ implicitly depend on temperature and chemical potential through the Fermi functions and phonon occupation number ($N_{\vex{q}}$) in Eq.~(\ref{RelaxationLengthSelfConsistency}). Once the $\{l_{\vex{k}}\}$ are known for a given pair $(n,T)$, the resistivity can be computed through Eq.~(\ref{ResistivityDefinition}). 

It is important to note that as we scan $T$ for a fixed $n$, $\mu(n,T)$ can change, and this can be quite drastic near a gap, a Van Hove singularity, and especially in the presence of a flat band. We must therefore calculate $\mu(n,T)$ self-consistently via 
\begin{align}
    \label{SelfConsistentChemicalPotential}
    n = \frac{4}{\mathcal{A}}\sum_{S} f(\e).
\end{align}
The prefactor 4 above follows from the spin and valley degeneracies. We stress that accurately computing the $T$-dependence of $\mu(n,T)$ near band edges, Van-Hove singularities, or in nearly flat bands, requires keeping extensive bandstructure data in this calculation, even if $T$ is far too low to excite carriers in these features. This can be a non-trivial numerical task, requiring great accuracy. 

The main result of this work is the application of the above analysis to twisted bilayer graphene at several twist angles near the ``magic" angle corresponding to the flat band. These results are presented and discussed already in Sec.~\ref{Section-SummaryOfMainResults}. We do this using the full BM Hamiltonian \cite{Bistritzer_2011}. The nontrivial band geometry of these systems gives scattering manifolds that depend qualitatively on not only the Fermi level, but also the specific Bloch state in question, as depicted in Fig.~\ref{IntroductionDoSFigure}. Since the bands are not isotropic and the phonon scattering cannot be considered ``quasi-elastic" \cite{SankarGrapheneKT1}, we need to find the full solution of Eq.~(\ref{RelaxationLengthSelfConsistency}). Solving Eq.~(\ref{RelaxationLengthSelfConsistency}) for the $\{l_k\}$ repeatedly for many values of $n$ and $T$, we calculate the resistivity data given in Figs.~\ref{LogRhoTFigure}-\ref{SecondarySlicesFigure}. 





\section{Discussion and conclusions}
\label{Section-DiscussionAndConclusions}

We have calculated the electrical resistivity of twisted bilayer graphene due to scattering from acoustic phonons. We extend previous studies by using the detailed BM band structure and focusing our attention on the effects due to the geometry of the band structure, including anisotropy, band curvature, excitation gaps, nearly flat bands, and Umklapp scattering across the moir\'{e} Brillouin zone.

We develop a thoroughly nontrivial transport theory for carrier resistivity due to electron-acoustic phonon interaction in twisted bilayer graphene, currently one of the most experimentally relevant systems to condensed matter physics. While we use the standard graphene acoustic phonons and the conventional electron-phonon deformation potential coupling, we include the full effects of the BM band structure. We use an iterative numerical technique to accurately solve the integral Boltzmann transport equation, and resolve the full effect of the BM band structure on the resistivity. This technique is able to incorporate the full complexity of the BM band structure, including the nearly flat bands, the Van Hove singularities, the curvature near the band edges, the anisotropy of the Fermi surface, and the abundance of Umklapp scattering due to the small moir\'{e} Brillouin zone. Inclusion of these geometric features leads to several new qualitative elements in the temperature and doping dependence of the resistivity, unlike those discussed in the transport literature (e.g. resistance peaks and apparent resistance saturation, nonlinear temperature dependence at high-$T$, complication of the BG crossover, among other features). We give concrete predictions for the doping and temperature dependence of the resistivity in TBLG, demonstrating how the results differ from the Dirac cone approximation. These results constitute a specific self-consistent theory to which experiments may be compared.

Our results are an important contribution to the ongoing discussion on the existence and nature of a strange metal phase in TBLG. It is crucial in the investigation of the origin of the superconductivity and presence of a strange metal phase in TBLG to understand the relative importance of various scattering mechanisms. Our work provides a clear and concrete picture of how the resistivity should behave in a phonon-dominated system. If strong deviations from these results are seen in experiment, that could serve as evidence that scattering mechanisms other than phonons may sometimes play dominant roles in transport. We emphasize that phonon scattering is always present, and must be considered in any discussion of the temperature dependent resistivity in any material, including TBLG. More generally, the exotic physics of TBLG is but one example of the capability of 2D layered heterostructures to host a wide variety of exotic phases and phenomena. As this class of materials has rapidly become an important subfield of condensed matter physics, it is imperative to study the relationship between their band geometries and transport directly.

Our results indicate that the BG crossover, below which the linear-in-T resistivity behavior is suppressed, begins - at all dopings and twist angles under study - at temperatures between $5K$ and $15K$, with an exception for fillings very close to the Dirac point. However, the width of the crossover region can change dramatically with doping, as the geometry of the Fermi surface changes. In particular, we see that as we dope out of the $|\nu| \leq 4$ regime, the $T^{\alpha > 1}$ scaling survives to much higher $T$. Further, we see that effects of curvature in the band, and low velocities in the nearly flat conduction/valence bands, cause regions of nonlinear $T$-dependence characterized by resistivity peaks which are often followed by downturns. These effects are the strongest in the $|\nu| \leq 4$ regime. 

Although our model is conceptually simple, only using basic kinetic theory principles, the single particle BM band structure, and the Debye approximation for phonons, our results are qualitatively and semiquantitatively compatible with the experiments. For $|\nu| \leq 4$, we find a resistivity that varies by several orders of magnitude as the twist angle is varied by degrees. We find consistently a low BG crossover temperature in the $5-15K$ range, and regions of rough linearity in the $10-50K$ regime. Above this, we find robust resistivity peaks located around $50-100K$, which are sharper for angles closer to the magic angle condition. At higher temperatures, we see a sharp decrease of the resistivity down to much more universal behavior that does not depend as sensitively on twist angle or doping. These are all consistent with the experimental TBLG data, and are not captured within the Dirac cone approximation.  In particular, the sharp increase in the resistivity at smaller twist angles (without any adjustment of the deformation potential coupling) and the resistivity peaks at higher temperatures are the qualitative hallmarks of our theory which have not been captured in the earlier theories of TBLG transport.  The linearity in temperature persisting down to $5K$ is also a new result in our theory, indicating that the linear-in-T resistivity in TBLG can persist to temperatures much lower than the nominal BG temperature of regular graphene.

Many features of our results follow from the nontrivial band geometry, and can not be predicted using the simple Dirac cone approximation. However, even at low temperatures, we see that Dirac cone approximation can significantly understate the resistivity when compared to the results of the full BM calculation. We compare the results of the two approaches extensively in Fig.~\ref{ComparisonWithWuDasSarmaFigure}. While the Dirac cone approximation captures the asymptotic low-$T$ physics for points very close to the Dirac point, for small twist angles we see that the full band structure leads to a nontrivial alteration of the resistivity. In fact, the enhancement of the resistivity we observe beyond the prediction of the Dirac cone approximation brings our theory's predictions much closer to the experimental observations \cite{Polshyn2019}.

We emphasize that the above results quite accurately capture the experimental picture for temperatures above $5K$. In particular, our calculation is able to not only explain the resistivity peaks observed in TBLG experiments, but also give quantitatively accurate estimates of the temperatures at which these peaks occur.

On the other hand, our results indicate that in the simplest phonon scattering picture, the BG crossover should suppress any linear scaling in resistivity with temperature at temperatures lower than $5K$ [Fig.~(\ref{TLinearFigure})]. Our results thus cannot explain the occasional reported observations of the ``strange metal" resistivity scaling at very low $T$ for twist angles in the $1.1^{\circ}-1.4^{\circ}$ range \cite{Jaoui_2022}.

We emphasize that myriad other mechanisms at play in TBLG can alter the resistivity from our predictions. For example, our calculation assumes that both the phonon band structure and the electron-phonon couplings are unchanged from their values in single-layer graphene and are unaffected by the moir\'{e} structure. We believe that these are reasonable approximations, but we cannot rule out the possibility that the phonon dispersion and the deformation potential coupling are somehow modified by the moir\'{e} TBLG structure, which may be responsible (e.g. a suppression of the phonon velocity) for the linearity in temperature to sometimes persist to very low temperatures.  In addition, at small twist angles, the Fermi velocity is comparable to the phonon velocity, which may lead to new flat band physics not included in our theory which applies only for the situation where the Fermi velocity is greater than the phonon velocity \cite{Davis_2023Subsonic}. 

Finally, our work focuses only on the electron-phonon interaction and neglects effects due to the electron-electron interaction. It is well-known that, due to Galilean invariance, electron-electron collisions only contribute to long wavelength resistivity via Umklapp processes, which suppresses their effect on transport in normal metals. This is simply because the center of mass and relative motions separate in the absence of Umklapp, and is formalized in the Prange-Kadanoff theorem \cite{Prange_1964, Prange_1967}. (For example, in normal metals the effective mass entering the Drude formula is the bare band mass and should not include any Fermi liquid renormalization.) However, since in TBLG the moir\'{e}-Brillouin zone is orders of magnitude smaller, Umklapp effects are more likely to be important. Additionally, the large density of states in the nearly flat bands could allow for very efficient screening that further reduces the importance of electron-electron collisions. The open questions of the importance of screening and moir\'{e}-Umklapp processes in electron-electron interactions in TBLG leave it ambiguous to what extent band renormalizations should be taken into account in a transport theory. It is, however, important to emphasize that the Umklapp electron-electron scattering traditionally leads to a $T^2$ resistivity generically, which is never observed in TBLG for $T > 5K$, \cite{Jaoui_2022} which is the regime of interest on our work (i.e. the equipartition temperate regime where electron-phonon scattering contributes substantially to the resistivity), providing some evidence for the absence of any direct electron interaction effects on the TBLG transport.

In particular, the Hartree effects arising from the electron-electron interactions can be included straightforwardly in our formalism if necessary. However, given the huge sample to sample quantitative differences in the experimentally reported TBLG resistivity, we believe that such a theory adds little to the current understanding of the TBLG transport properties where at this point only a qualitative and semi-quantitative theory based on the standard BM band structure model (as we do) is meaningful.  No theory can explain experimental results quantitatively when the experiments themselves show considerable variations. 

On the other hand, there is ample evidence \cite{Choi_2021_InteractionDriven, Lewandowski_2021_FillingDependent, Rademaker_2019} that interaction effects lead to a strongly doping-dependent renormalization of the nearly flat bands from their non-interacting form as well as to so-called ``cascade physics" \cite{Wong_2020, Zondiner_2020} that cause spin-valley polarization in the ground state. In principle, these effects alter the effective band structure and including them in our calculation would lead to quantitative alterations of our transport theory. This could be a productive direction for future work. Conversely, our transport theory of non-interacting electrons provide a benchmark that can be compared with experiment to probe the extent to which interaction effects indeed modify transport. In particular, our results show that the temperature at which the resistivity peaks is largely insensitive to the doping throughout the nearly flat band [Figs.~\ref{PrimarySlicesFigure}-\ref{SecondarySlicesFigure}], and that the temperature at which the peak is found is roughly correlated to the bandwidth of the nearly flat band. If experimental transport data were to consistently show that the temperature of the resistivity peak shifts strongly with doping, this would be clear evidence of interaction-induced renormalization indirectly contributing to transport. Our results may thus be combined with a detailed experimental analysis of the dependence of the resistivity peak temperatures on doping to understand the importance of electron-electron renormalizations in TBLG transport.




\acknowledgments

We thank Yang-Zhi Chou, Matthew S. Foster, Christopher D. White, Jiabin Yu, Dabanjan Chowdhury, and Allan MacDonald for helpful discussions. This work is supported by the Laboratory for Physical Sciences (S.M.D, and S.D.S). F. W. is supported by National Natural Science Foundation of China (Grant No. 12274333) and start-up funding of Wuhan University.

\appendix

\bibliography{BIB}

\begin{thebibliography}{67}
\expandafter\ifx\csname natexlab\endcsname\relax\def\natexlab#1{#1}\fi
\expandafter\ifx\csname bibnamefont\endcsname\relax
  \def\bibnamefont#1{#1}\fi
\expandafter\ifx\csname bibfnamefont\endcsname\relax
  \def\bibfnamefont#1{#1}\fi
\expandafter\ifx\csname citenamefont\endcsname\relax
  \def\citenamefont#1{#1}\fi
\expandafter\ifx\csname url\endcsname\relax
  \def\url#1{\texttt{#1}}\fi
\expandafter\ifx\csname urlprefix\endcsname\relax\def\urlprefix{URL }\fi
\providecommand{\bibinfo}[2]{#2}
\providecommand{\eprint}[2][]{\url{#2}}

\bibitem[{\citenamefont{Cao et~al.}(2018{\natexlab{a}})\citenamefont{Cao,
  Fatemi, Demir, Fang, Tomarken, Luo, Sanchez-Yamagishi, Watanabe, Taniguchi,
  Kaxiras et~al.}}]{Cao_2018a}
\bibinfo{author}{\bibfnamefont{Y.}~\bibnamefont{Cao}},
  \bibinfo{author}{\bibfnamefont{V.}~\bibnamefont{Fatemi}},
  \bibinfo{author}{\bibfnamefont{A.}~\bibnamefont{Demir}},
  \bibinfo{author}{\bibfnamefont{S.}~\bibnamefont{Fang}},
  \bibinfo{author}{\bibfnamefont{S.~L.} \bibnamefont{Tomarken}},
  \bibinfo{author}{\bibfnamefont{J.~Y.} \bibnamefont{Luo}},
  \bibinfo{author}{\bibfnamefont{J.~D.} \bibnamefont{Sanchez-Yamagishi}},
  \bibinfo{author}{\bibfnamefont{K.}~\bibnamefont{Watanabe}},
  \bibinfo{author}{\bibfnamefont{T.}~\bibnamefont{Taniguchi}},
  \bibinfo{author}{\bibfnamefont{E.}~\bibnamefont{Kaxiras}},
  \bibnamefont{et~al.}, \bibinfo{journal}{Nature}
  \textbf{\bibinfo{volume}{556}}, \bibinfo{pages}{80}
  (\bibinfo{year}{2018}{\natexlab{a}}),
  \urlprefix\url{https://doi.org/10.1038%2Fnature26154}.

\bibitem[{\citenamefont{Cao et~al.}(2018{\natexlab{b}})\citenamefont{Cao,
  Fatemi, Fang, Watanabe, Taniguchi, Kaxiras, and Jarillo-Herrero}}]{Cao_2018b}
\bibinfo{author}{\bibfnamefont{Y.}~\bibnamefont{Cao}},
  \bibinfo{author}{\bibfnamefont{V.}~\bibnamefont{Fatemi}},
  \bibinfo{author}{\bibfnamefont{S.}~\bibnamefont{Fang}},
  \bibinfo{author}{\bibfnamefont{K.}~\bibnamefont{Watanabe}},
  \bibinfo{author}{\bibfnamefont{T.}~\bibnamefont{Taniguchi}},
  \bibinfo{author}{\bibfnamefont{E.}~\bibnamefont{Kaxiras}}, \bibnamefont{and}
  \bibinfo{author}{\bibfnamefont{P.}~\bibnamefont{Jarillo-Herrero}},
  \bibinfo{journal}{Nature} \textbf{\bibinfo{volume}{556}}, \bibinfo{pages}{43}
  (\bibinfo{year}{2018}{\natexlab{b}}),
  \urlprefix\url{https://doi.org/10.1038%2Fnature26160}.

\bibitem[{\citenamefont{Cao et~al.}(2020{\natexlab{a}})\citenamefont{Cao,
  Chowdhury, Rodan-Legrain, Rubies-Bigorda, Watanabe, Taniguchi, Senthil, and
  Jarillo-Herrero}}]{Cao2020PRL}
\bibinfo{author}{\bibfnamefont{Y.}~\bibnamefont{Cao}},
  \bibinfo{author}{\bibfnamefont{D.}~\bibnamefont{Chowdhury}},
  \bibinfo{author}{\bibfnamefont{D.}~\bibnamefont{Rodan-Legrain}},
  \bibinfo{author}{\bibfnamefont{O.}~\bibnamefont{Rubies-Bigorda}},
  \bibinfo{author}{\bibfnamefont{K.}~\bibnamefont{Watanabe}},
  \bibinfo{author}{\bibfnamefont{T.}~\bibnamefont{Taniguchi}},
  \bibinfo{author}{\bibfnamefont{T.}~\bibnamefont{Senthil}}, \bibnamefont{and}
  \bibinfo{author}{\bibfnamefont{P.}~\bibnamefont{Jarillo-Herrero}},
  \bibinfo{journal}{Phys. Rev. Lett.} \textbf{\bibinfo{volume}{124}},
  \bibinfo{pages}{076801} (\bibinfo{year}{2020}{\natexlab{a}}),
  \urlprefix\url{https://link.aps.org/doi/10.1103/PhysRevLett.124.076801}.

\bibitem[{\citenamefont{Cao et~al.}(2021)\citenamefont{Cao, Park, Watanabe,
  Taniguchi, and Jarillo-Herrero}}]{Cao2021}
\bibinfo{author}{\bibfnamefont{Y.}~\bibnamefont{Cao}},
  \bibinfo{author}{\bibfnamefont{J.~M.} \bibnamefont{Park}},
  \bibinfo{author}{\bibfnamefont{K.}~\bibnamefont{Watanabe}},
  \bibinfo{author}{\bibfnamefont{T.}~\bibnamefont{Taniguchi}},
  \bibnamefont{and}
  \bibinfo{author}{\bibfnamefont{P.}~\bibnamefont{Jarillo-Herrero}},
  \bibinfo{journal}{Nature} \textbf{\bibinfo{volume}{595}},
  \bibinfo{pages}{526} (\bibinfo{year}{2021}).

\bibitem[{\citenamefont{Yankowitz et~al.}(2019)\citenamefont{Yankowitz, Chen,
  Polshyn, Zhang, Watanabe, Taniguchi, Graf, Young, and Dean}}]{Yankowitz_2019}
\bibinfo{author}{\bibfnamefont{M.}~\bibnamefont{Yankowitz}},
  \bibinfo{author}{\bibfnamefont{S.}~\bibnamefont{Chen}},
  \bibinfo{author}{\bibfnamefont{H.}~\bibnamefont{Polshyn}},
  \bibinfo{author}{\bibfnamefont{Y.}~\bibnamefont{Zhang}},
  \bibinfo{author}{\bibfnamefont{K.}~\bibnamefont{Watanabe}},
  \bibinfo{author}{\bibfnamefont{T.}~\bibnamefont{Taniguchi}},
  \bibinfo{author}{\bibfnamefont{D.}~\bibnamefont{Graf}},
  \bibinfo{author}{\bibfnamefont{A.~F.} \bibnamefont{Young}}, \bibnamefont{and}
  \bibinfo{author}{\bibfnamefont{C.~R.} \bibnamefont{Dean}},
  \bibinfo{journal}{Science} \textbf{\bibinfo{volume}{363}},
  \bibinfo{pages}{1059} (\bibinfo{year}{2019}),
  \urlprefix\url{https://doi.org/10.1126%2Fscience.aav1910}.

\bibitem[{\citenamefont{Kerelsky et~al.}(2019)\citenamefont{Kerelsky, McGilly,
  Kennes, Xian, Yankowitz, Chen, Watanabe, Taniguchi, Hone, Dean
  et~al.}}]{Kerelsky_2019}
\bibinfo{author}{\bibfnamefont{A.}~\bibnamefont{Kerelsky}},
  \bibinfo{author}{\bibfnamefont{L.~J.} \bibnamefont{McGilly}},
  \bibinfo{author}{\bibfnamefont{D.~M.} \bibnamefont{Kennes}},
  \bibinfo{author}{\bibfnamefont{L.}~\bibnamefont{Xian}},
  \bibinfo{author}{\bibfnamefont{M.}~\bibnamefont{Yankowitz}},
  \bibinfo{author}{\bibfnamefont{S.}~\bibnamefont{Chen}},
  \bibinfo{author}{\bibfnamefont{K.}~\bibnamefont{Watanabe}},
  \bibinfo{author}{\bibfnamefont{T.}~\bibnamefont{Taniguchi}},
  \bibinfo{author}{\bibfnamefont{J.}~\bibnamefont{Hone}},
  \bibinfo{author}{\bibfnamefont{C.}~\bibnamefont{Dean}}, \bibnamefont{et~al.},
  \bibinfo{journal}{Nature} \textbf{\bibinfo{volume}{572}}, \bibinfo{pages}{95}
  (\bibinfo{year}{2019}),
  \urlprefix\url{https://doi.org/10.1038%2Fs41586-019-1431-9}.

\bibitem[{\citenamefont{Lu et~al.}(2019)\citenamefont{Lu, Stepanov, Yang, Xie,
  Aamir, Das, Urgell, Watanabe, Taniguchi, Zhang et~al.}}]{Lu_2019}
\bibinfo{author}{\bibfnamefont{X.}~\bibnamefont{Lu}},
  \bibinfo{author}{\bibfnamefont{P.}~\bibnamefont{Stepanov}},
  \bibinfo{author}{\bibfnamefont{W.}~\bibnamefont{Yang}},
  \bibinfo{author}{\bibfnamefont{M.}~\bibnamefont{Xie}},
  \bibinfo{author}{\bibfnamefont{M.~A.} \bibnamefont{Aamir}},
  \bibinfo{author}{\bibfnamefont{I.}~\bibnamefont{Das}},
  \bibinfo{author}{\bibfnamefont{C.}~\bibnamefont{Urgell}},
  \bibinfo{author}{\bibfnamefont{K.}~\bibnamefont{Watanabe}},
  \bibinfo{author}{\bibfnamefont{T.}~\bibnamefont{Taniguchi}},
  \bibinfo{author}{\bibfnamefont{G.}~\bibnamefont{Zhang}},
  \bibnamefont{et~al.}, \bibinfo{journal}{Nature}
  \textbf{\bibinfo{volume}{574}}, \bibinfo{pages}{653} (\bibinfo{year}{2019}),
  \urlprefix\url{https://doi.org/10.1038%2Fs41586-019-1695-0}.

\bibitem[{\citenamefont{Wu et~al.}(2018)\citenamefont{Wu, MacDonald, and
  Martin}}]{Wu_2018}
\bibinfo{author}{\bibfnamefont{F.}~\bibnamefont{Wu}},
  \bibinfo{author}{\bibfnamefont{A.}~\bibnamefont{MacDonald}},
  \bibnamefont{and} \bibinfo{author}{\bibfnamefont{I.}~\bibnamefont{Martin}},
  \bibinfo{journal}{Physical Review Letters} \textbf{\bibinfo{volume}{121}}
  (\bibinfo{year}{2018}),
  \urlprefix\url{https://doi.org/10.1103%2Fphysrevlett.121.257001}.

\bibitem[{\citenamefont{Bistritzer and MacDonald}(2011)}]{Bistritzer_2011}
\bibinfo{author}{\bibfnamefont{R.}~\bibnamefont{Bistritzer}} \bibnamefont{and}
  \bibinfo{author}{\bibfnamefont{A.~H.} \bibnamefont{MacDonald}},
  \bibinfo{journal}{Proceedings of the National Academy of Sciences}
  \textbf{\bibinfo{volume}{108}}, \bibinfo{pages}{12233}
  (\bibinfo{year}{2011}),
  \urlprefix\url{https://doi.org/10.1073%2Fpnas.1108174108}.

\bibitem[{\citenamefont{Wu et~al.}(2019{\natexlab{a}})\citenamefont{Wu, Hwang,
  and Das~Sarma}}]{Wu2019_phonon}
\bibinfo{author}{\bibfnamefont{F.}~\bibnamefont{Wu}},
  \bibinfo{author}{\bibfnamefont{E.}~\bibnamefont{Hwang}}, \bibnamefont{and}
  \bibinfo{author}{\bibfnamefont{S.}~\bibnamefont{Das~Sarma}},
  \bibinfo{journal}{Phys. Rev. B} \textbf{\bibinfo{volume}{99}},
  \bibinfo{pages}{165112} (\bibinfo{year}{2019}{\natexlab{a}}),
  \urlprefix\url{https://link.aps.org/doi/10.1103/PhysRevB.99.165112}.

\bibitem[{\citenamefont{Li et~al.}(2020)\citenamefont{Li, Wu, and
  Das~Sarma}}]{Li2020}
\bibinfo{author}{\bibfnamefont{X.}~\bibnamefont{Li}},
  \bibinfo{author}{\bibfnamefont{F.}~\bibnamefont{Wu}}, \bibnamefont{and}
  \bibinfo{author}{\bibfnamefont{S.}~\bibnamefont{Das~Sarma}},
  \bibinfo{journal}{Phys. Rev. B} \textbf{\bibinfo{volume}{101}},
  \bibinfo{pages}{245436} (\bibinfo{year}{2020}),
  \urlprefix\url{https://link.aps.org/doi/10.1103/PhysRevB.101.245436}.

\bibitem[{\citenamefont{Stepanov et~al.}(2020)\citenamefont{Stepanov, Das, Lu,
  Fahimniya, Watanabe, Taniguchi, Koppens, Lischner, Levitov, and
  Efetov}}]{Stepanov2020untying}
\bibinfo{author}{\bibfnamefont{P.}~\bibnamefont{Stepanov}},
  \bibinfo{author}{\bibfnamefont{I.}~\bibnamefont{Das}},
  \bibinfo{author}{\bibfnamefont{X.}~\bibnamefont{Lu}},
  \bibinfo{author}{\bibfnamefont{A.}~\bibnamefont{Fahimniya}},
  \bibinfo{author}{\bibfnamefont{K.}~\bibnamefont{Watanabe}},
  \bibinfo{author}{\bibfnamefont{T.}~\bibnamefont{Taniguchi}},
  \bibinfo{author}{\bibfnamefont{F.~H.} \bibnamefont{Koppens}},
  \bibinfo{author}{\bibfnamefont{J.}~\bibnamefont{Lischner}},
  \bibinfo{author}{\bibfnamefont{L.}~\bibnamefont{Levitov}}, \bibnamefont{and}
  \bibinfo{author}{\bibfnamefont{D.~K.} \bibnamefont{Efetov}},
  \bibinfo{journal}{Nature} \textbf{\bibinfo{volume}{583}},
  \bibinfo{pages}{375} (\bibinfo{year}{2020}).

\bibitem[{\citenamefont{Saito et~al.}(2020)\citenamefont{Saito, Ge, Watanabe,
  Taniguchi, and Young}}]{Saito2020independent}
\bibinfo{author}{\bibfnamefont{Y.}~\bibnamefont{Saito}},
  \bibinfo{author}{\bibfnamefont{J.}~\bibnamefont{Ge}},
  \bibinfo{author}{\bibfnamefont{K.}~\bibnamefont{Watanabe}},
  \bibinfo{author}{\bibfnamefont{T.}~\bibnamefont{Taniguchi}},
  \bibnamefont{and} \bibinfo{author}{\bibfnamefont{A.~F.} \bibnamefont{Young}},
  \bibinfo{journal}{Nature Physics} \textbf{\bibinfo{volume}{16}},
  \bibinfo{pages}{926} (\bibinfo{year}{2020}).

\bibitem[{\citenamefont{Sarma and Wu}(2020)}]{Das_Sarma_2020}
\bibinfo{author}{\bibfnamefont{S.~D.} \bibnamefont{Sarma}} \bibnamefont{and}
  \bibinfo{author}{\bibfnamefont{F.}~\bibnamefont{Wu}},
  \bibinfo{journal}{Annals of Physics} \textbf{\bibinfo{volume}{417}},
  \bibinfo{pages}{168193} (\bibinfo{year}{2020}),
  \urlprefix\url{https://doi.org/10.1016%2Fj.aop.2020.168193}.

\bibitem[{\citenamefont{Jaoui et~al.}(2021)\citenamefont{Jaoui, Das,
  Di~Battista, Díez-Mérida, Lu, Watanabe, Taniguchi, Ishizuka, Levitov, and
  Efetov}}]{TBLGStrangeMetalExperiment1}
\bibinfo{author}{\bibfnamefont{A.}~\bibnamefont{Jaoui}},
  \bibinfo{author}{\bibfnamefont{I.}~\bibnamefont{Das}},
  \bibinfo{author}{\bibfnamefont{G.}~\bibnamefont{Di~Battista}},
  \bibinfo{author}{\bibfnamefont{J.}~\bibnamefont{Díez-Mérida}},
  \bibinfo{author}{\bibfnamefont{X.}~\bibnamefont{Lu}},
  \bibinfo{author}{\bibfnamefont{K.}~\bibnamefont{Watanabe}},
  \bibinfo{author}{\bibfnamefont{T.}~\bibnamefont{Taniguchi}},
  \bibinfo{author}{\bibfnamefont{H.}~\bibnamefont{Ishizuka}},
  \bibinfo{author}{\bibfnamefont{L.}~\bibnamefont{Levitov}}, \bibnamefont{and}
  \bibinfo{author}{\bibfnamefont{D.~K.} \bibnamefont{Efetov}},
  \emph{\bibinfo{title}{Quantum critical behavior in magic-angle twisted
  bilayer graphene}} (\bibinfo{year}{2021}),
  \urlprefix\url{https://arxiv.org/abs/2108.07753}.

\bibitem[{\citenamefont{Polshyn et~al.}(2019)\citenamefont{Polshyn, Yankowitz,
  Chen, Zhang, Watanabe, Taniguchi, Dean, and Young}}]{Polshyn2019}
\bibinfo{author}{\bibfnamefont{H.}~\bibnamefont{Polshyn}},
  \bibinfo{author}{\bibfnamefont{M.}~\bibnamefont{Yankowitz}},
  \bibinfo{author}{\bibfnamefont{S.}~\bibnamefont{Chen}},
  \bibinfo{author}{\bibfnamefont{Y.}~\bibnamefont{Zhang}},
  \bibinfo{author}{\bibfnamefont{K.}~\bibnamefont{Watanabe}},
  \bibinfo{author}{\bibfnamefont{T.}~\bibnamefont{Taniguchi}},
  \bibinfo{author}{\bibfnamefont{C.~R.} \bibnamefont{Dean}}, \bibnamefont{and}
  \bibinfo{author}{\bibfnamefont{A.~F.} \bibnamefont{Young}},
  \bibinfo{journal}{Nature Physics} \textbf{\bibinfo{volume}{15}},
  \bibinfo{pages}{1011} (\bibinfo{year}{2019}).

\bibitem[{\citenamefont{Cao et~al.}(2020{\natexlab{b}})\citenamefont{Cao,
  Chowdhury, Rodan-Legrain, Rubies-Bigorda, Watanabe, Taniguchi, Senthil, and
  Jarillo-Herrero}}]{TBLGStrangeMetalExperiment3}
\bibinfo{author}{\bibfnamefont{Y.}~\bibnamefont{Cao}},
  \bibinfo{author}{\bibfnamefont{D.}~\bibnamefont{Chowdhury}},
  \bibinfo{author}{\bibfnamefont{D.}~\bibnamefont{Rodan-Legrain}},
  \bibinfo{author}{\bibfnamefont{O.}~\bibnamefont{Rubies-Bigorda}},
  \bibinfo{author}{\bibfnamefont{K.}~\bibnamefont{Watanabe}},
  \bibinfo{author}{\bibfnamefont{T.}~\bibnamefont{Taniguchi}},
  \bibinfo{author}{\bibfnamefont{T.}~\bibnamefont{Senthil}}, \bibnamefont{and}
  \bibinfo{author}{\bibfnamefont{P.}~\bibnamefont{Jarillo-Herrero}},
  \bibinfo{journal}{Phys. Rev. Lett.} \textbf{\bibinfo{volume}{124}},
  \bibinfo{pages}{076801} (\bibinfo{year}{2020}{\natexlab{b}}),
  \urlprefix\url{https://link.aps.org/doi/10.1103/PhysRevLett.124.076801}.

\bibitem[{\citenamefont{Sarma and Wu}(2022)}]{SankarFengchengStrangeMetal}
\bibinfo{author}{\bibfnamefont{S.~D.} \bibnamefont{Sarma}} \bibnamefont{and}
  \bibinfo{author}{\bibfnamefont{F.}~\bibnamefont{Wu}},
  \bibinfo{journal}{Physical Review Research} \textbf{\bibinfo{volume}{4}}
  (\bibinfo{year}{2022}),
  \urlprefix\url{https://doi.org/10.1103%2Fphysrevresearch.4.033061}.

\bibitem[{\citenamefont{Hwang and Sarma}(2019{\natexlab{a}})}]{Hwang_2019}
\bibinfo{author}{\bibfnamefont{E.~H.} \bibnamefont{Hwang}} \bibnamefont{and}
  \bibinfo{author}{\bibfnamefont{S.~D.} \bibnamefont{Sarma}},
  \bibinfo{journal}{Physical Review B} \textbf{\bibinfo{volume}{99}}
  (\bibinfo{year}{2019}{\natexlab{a}}),
  \urlprefix\url{https://doi.org/10.1103%2Fphysrevb.99.085105}.

\bibitem[{\citenamefont{Ashcroft and Mermin}(1976)}]{AshcroftAndMermin}
\bibinfo{author}{\bibfnamefont{N.~W.} \bibnamefont{Ashcroft}} \bibnamefont{and}
  \bibinfo{author}{\bibfnamefont{N.~D.} \bibnamefont{Mermin}},
  \emph{\bibinfo{title}{Solid State Physics}} (\bibinfo{publisher}{Harcourt
  College Publishers}, \bibinfo{year}{1976}), ISBN
  \bibinfo{isbn}{9780030839931}.

\bibitem[{\citenamefont{Hwang and Sarma}(2008)}]{SankarGrapheneKT1}
\bibinfo{author}{\bibfnamefont{E.~H.} \bibnamefont{Hwang}} \bibnamefont{and}
  \bibinfo{author}{\bibfnamefont{S.~D.} \bibnamefont{Sarma}},
  \bibinfo{journal}{Physical Review B} \textbf{\bibinfo{volume}{77}}
  (\bibinfo{year}{2008}),
  \urlprefix\url{https://doi.org/10.1103%2Fphysrevb.77.115449}.

\bibitem[{\citenamefont{Hwang and Das~Sarma}(2008)}]{Hwang2008}
\bibinfo{author}{\bibfnamefont{E.~H.} \bibnamefont{Hwang}} \bibnamefont{and}
  \bibinfo{author}{\bibfnamefont{S.}~\bibnamefont{Das~Sarma}},
  \bibinfo{journal}{Phys. Rev. B} \textbf{\bibinfo{volume}{77}},
  \bibinfo{pages}{115449} (\bibinfo{year}{2008}),
  \urlprefix\url{https://link.aps.org/doi/10.1103/PhysRevB.77.115449}.

\bibitem[{\citenamefont{Min et~al.}(2011{\natexlab{a}})\citenamefont{Min,
  Hwang, and Das~Sarma}}]{Min2011}
\bibinfo{author}{\bibfnamefont{H.}~\bibnamefont{Min}},
  \bibinfo{author}{\bibfnamefont{E.~H.} \bibnamefont{Hwang}}, \bibnamefont{and}
  \bibinfo{author}{\bibfnamefont{S.}~\bibnamefont{Das~Sarma}},
  \bibinfo{journal}{Phys. Rev. B} \textbf{\bibinfo{volume}{83}},
  \bibinfo{pages}{161404} (\bibinfo{year}{2011}{\natexlab{a}}),
  \urlprefix\url{https://link.aps.org/doi/10.1103/PhysRevB.83.161404}.

\bibitem[{\citenamefont{Efetov and Kim}(2010)}]{Efetov_2010}
\bibinfo{author}{\bibfnamefont{D.~K.} \bibnamefont{Efetov}} \bibnamefont{and}
  \bibinfo{author}{\bibfnamefont{P.}~\bibnamefont{Kim}},
  \bibinfo{journal}{Physical Review Letters} \textbf{\bibinfo{volume}{105}}
  (\bibinfo{year}{2010}),
  \urlprefix\url{https://doi.org/10.1103%2Fphysrevlett.105.256805}.

\bibitem[{\citenamefont{Min et~al.}(2011{\natexlab{b}})\citenamefont{Min,
  Hwang, and Sarma}}]{SankarGrapheneKT2}
\bibinfo{author}{\bibfnamefont{H.}~\bibnamefont{Min}},
  \bibinfo{author}{\bibfnamefont{E.~H.} \bibnamefont{Hwang}}, \bibnamefont{and}
  \bibinfo{author}{\bibfnamefont{S.~D.} \bibnamefont{Sarma}},
  \bibinfo{journal}{Physical Review B} \textbf{\bibinfo{volume}{83}}
  (\bibinfo{year}{2011}{\natexlab{b}}),
  \urlprefix\url{https://doi.org/10.1103%2Fphysrevb.83.161404}.

\bibitem[{\citenamefont{Hwang and
  Sarma}(2019{\natexlab{b}})}]{SankarGrapheneKT5}
\bibinfo{author}{\bibfnamefont{E.~H.} \bibnamefont{Hwang}} \bibnamefont{and}
  \bibinfo{author}{\bibfnamefont{S.~D.} \bibnamefont{Sarma}},
  \bibinfo{journal}{Physical Review B} \textbf{\bibinfo{volume}{99}}
  (\bibinfo{year}{2019}{\natexlab{b}}),
  \urlprefix\url{https://doi.org/10.1103%2Fphysrevb.99.085105}.

\bibitem[{\citenamefont{Ziman}(1960)}]{Ziman}
\bibinfo{author}{\bibfnamefont{J.~M.} \bibnamefont{Ziman}},
  \emph{\bibinfo{title}{Electrons and Phonons}} (\bibinfo{publisher}{Oxford
  University Press}, \bibinfo{year}{1960}), ISBN \bibinfo{isbn}{9780198507796}.

\bibitem[{\citenamefont{Geim and Grigorieva}(2013)}]{Geim_2013}
\bibinfo{author}{\bibfnamefont{A.~K.} \bibnamefont{Geim}} \bibnamefont{and}
  \bibinfo{author}{\bibfnamefont{I.~V.} \bibnamefont{Grigorieva}},
  \bibinfo{journal}{Nature} \textbf{\bibinfo{volume}{499}},
  \bibinfo{pages}{419} (\bibinfo{year}{2013}),
  \urlprefix\url{https://doi.org/10.1038%2Fnature12385}.

\bibitem[{\citenamefont{Novoselov et~al.}(2006)\citenamefont{Novoselov, McCann,
  Morozov, Fal'ko, Katsnelson, Zeitler, Jiang, Schedin, and
  Geim}}]{Novoselov_2006}
\bibinfo{author}{\bibfnamefont{K.~S.} \bibnamefont{Novoselov}},
  \bibinfo{author}{\bibfnamefont{E.}~\bibnamefont{McCann}},
  \bibinfo{author}{\bibfnamefont{S.~V.} \bibnamefont{Morozov}},
  \bibinfo{author}{\bibfnamefont{V.~I.} \bibnamefont{Fal'ko}},
  \bibinfo{author}{\bibfnamefont{M.~I.} \bibnamefont{Katsnelson}},
  \bibinfo{author}{\bibfnamefont{U.}~\bibnamefont{Zeitler}},
  \bibinfo{author}{\bibfnamefont{D.}~\bibnamefont{Jiang}},
  \bibinfo{author}{\bibfnamefont{F.}~\bibnamefont{Schedin}}, \bibnamefont{and}
  \bibinfo{author}{\bibfnamefont{A.~K.} \bibnamefont{Geim}},
  \bibinfo{journal}{Nature Physics} \textbf{\bibinfo{volume}{2}},
  \bibinfo{pages}{177} (\bibinfo{year}{2006}),
  \urlprefix\url{https://doi.org/10.1038%2Fnphys245}.

\bibitem[{\citenamefont{Morell et~al.}(2010)\citenamefont{Morell, Correa,
  Vargas, Pacheco, and Barticevic}}]{Morell_2010}
\bibinfo{author}{\bibfnamefont{E.~S.} \bibnamefont{Morell}},
  \bibinfo{author}{\bibfnamefont{J.~D.} \bibnamefont{Correa}},
  \bibinfo{author}{\bibfnamefont{P.}~\bibnamefont{Vargas}},
  \bibinfo{author}{\bibfnamefont{M.}~\bibnamefont{Pacheco}}, \bibnamefont{and}
  \bibinfo{author}{\bibfnamefont{Z.}~\bibnamefont{Barticevic}},
  \bibinfo{journal}{Physical Review B} \textbf{\bibinfo{volume}{82}}
  (\bibinfo{year}{2010}),
  \urlprefix\url{https://doi.org/10.1103%2Fphysrevb.82.121407}.

\bibitem[{\citenamefont{Li et~al.}(2019)\citenamefont{Li, Wu, and
  MacDonald}}]{Li_2019}
\bibinfo{author}{\bibfnamefont{X.}~\bibnamefont{Li}},
  \bibinfo{author}{\bibfnamefont{F.}~\bibnamefont{Wu}}, \bibnamefont{and}
  \bibinfo{author}{\bibfnamefont{A.~H.} \bibnamefont{MacDonald}},
  \emph{\bibinfo{title}{Electronic structure of single-twist trilayer
  graphene}} (\bibinfo{year}{2019}),
  \urlprefix\url{https://arxiv.org/abs/1907.12338}.

\bibitem[{\citenamefont{Kim et~al.}(2017)\citenamefont{Kim, DaSilva, Huang,
  Fallahazad, Larentis, Taniguchi, Watanabe, LeRoy, MacDonald, and
  Tutuc}}]{Kim_2017}
\bibinfo{author}{\bibfnamefont{K.}~\bibnamefont{Kim}},
  \bibinfo{author}{\bibfnamefont{A.}~\bibnamefont{DaSilva}},
  \bibinfo{author}{\bibfnamefont{S.}~\bibnamefont{Huang}},
  \bibinfo{author}{\bibfnamefont{B.}~\bibnamefont{Fallahazad}},
  \bibinfo{author}{\bibfnamefont{S.}~\bibnamefont{Larentis}},
  \bibinfo{author}{\bibfnamefont{T.}~\bibnamefont{Taniguchi}},
  \bibinfo{author}{\bibfnamefont{K.}~\bibnamefont{Watanabe}},
  \bibinfo{author}{\bibfnamefont{B.~J.} \bibnamefont{LeRoy}},
  \bibinfo{author}{\bibfnamefont{A.~H.} \bibnamefont{MacDonald}},
  \bibnamefont{and} \bibinfo{author}{\bibfnamefont{E.}~\bibnamefont{Tutuc}},
  \bibinfo{journal}{Proceedings of the National Academy of Sciences}
  \textbf{\bibinfo{volume}{114}}, \bibinfo{pages}{3364} (\bibinfo{year}{2017}),
  \urlprefix\url{https://doi.org/10.1073%2Fpnas.1620140114}.

\bibitem[{\citenamefont{Sharpe et~al.}(2019)\citenamefont{Sharpe, Fox, Barnard,
  Finney, Watanabe, Taniguchi, Kastner, and Goldhaber-Gordon}}]{Sharpe_2019}
\bibinfo{author}{\bibfnamefont{A.~L.} \bibnamefont{Sharpe}},
  \bibinfo{author}{\bibfnamefont{E.~J.} \bibnamefont{Fox}},
  \bibinfo{author}{\bibfnamefont{A.~W.} \bibnamefont{Barnard}},
  \bibinfo{author}{\bibfnamefont{J.}~\bibnamefont{Finney}},
  \bibinfo{author}{\bibfnamefont{K.}~\bibnamefont{Watanabe}},
  \bibinfo{author}{\bibfnamefont{T.}~\bibnamefont{Taniguchi}},
  \bibinfo{author}{\bibfnamefont{M.~A.} \bibnamefont{Kastner}},
  \bibnamefont{and}
  \bibinfo{author}{\bibfnamefont{D.}~\bibnamefont{Goldhaber-Gordon}},
  \bibinfo{journal}{Science} \textbf{\bibinfo{volume}{365}},
  \bibinfo{pages}{605} (\bibinfo{year}{2019}),
  \urlprefix\url{https://doi.org/10.1126%2Fscience.aaw3780}.

\bibitem[{\citenamefont{Chen et~al.}(2020)\citenamefont{Chen, Sharpe, Fox,
  Zhang, Wang, Jiang, Lyu, Li, Watanabe, Taniguchi et~al.}}]{Chen_2020}
\bibinfo{author}{\bibfnamefont{G.}~\bibnamefont{Chen}},
  \bibinfo{author}{\bibfnamefont{A.~L.} \bibnamefont{Sharpe}},
  \bibinfo{author}{\bibfnamefont{E.~J.} \bibnamefont{Fox}},
  \bibinfo{author}{\bibfnamefont{Y.-H.} \bibnamefont{Zhang}},
  \bibinfo{author}{\bibfnamefont{S.}~\bibnamefont{Wang}},
  \bibinfo{author}{\bibfnamefont{L.}~\bibnamefont{Jiang}},
  \bibinfo{author}{\bibfnamefont{B.}~\bibnamefont{Lyu}},
  \bibinfo{author}{\bibfnamefont{H.}~\bibnamefont{Li}},
  \bibinfo{author}{\bibfnamefont{K.}~\bibnamefont{Watanabe}},
  \bibinfo{author}{\bibfnamefont{T.}~\bibnamefont{Taniguchi}},
  \bibnamefont{et~al.}, \bibinfo{journal}{Nature}
  \textbf{\bibinfo{volume}{579}}, \bibinfo{pages}{56} (\bibinfo{year}{2020}),
  \urlprefix\url{https://doi.org/10.1038%2Fs41586-020-2049-7}.

\bibitem[{\citenamefont{Rozen et~al.}(2021)\citenamefont{Rozen, Park, Zondiner,
  Cao, Rodan-Legrain, Taniguchi, Watanabe, Oreg, Stern, Berg
  et~al.}}]{Rozen2021entropic}
\bibinfo{author}{\bibfnamefont{A.}~\bibnamefont{Rozen}},
  \bibinfo{author}{\bibfnamefont{J.~M.} \bibnamefont{Park}},
  \bibinfo{author}{\bibfnamefont{U.}~\bibnamefont{Zondiner}},
  \bibinfo{author}{\bibfnamefont{Y.}~\bibnamefont{Cao}},
  \bibinfo{author}{\bibfnamefont{D.}~\bibnamefont{Rodan-Legrain}},
  \bibinfo{author}{\bibfnamefont{T.}~\bibnamefont{Taniguchi}},
  \bibinfo{author}{\bibfnamefont{K.}~\bibnamefont{Watanabe}},
  \bibinfo{author}{\bibfnamefont{Y.}~\bibnamefont{Oreg}},
  \bibinfo{author}{\bibfnamefont{A.}~\bibnamefont{Stern}},
  \bibinfo{author}{\bibfnamefont{E.}~\bibnamefont{Berg}}, \bibnamefont{et~al.},
  \bibinfo{journal}{Nature} \textbf{\bibinfo{volume}{592}},
  \bibinfo{pages}{214} (\bibinfo{year}{2021}).

\bibitem[{\citenamefont{Zhou et~al.}(2022)\citenamefont{Zhou, Holleis, Saito,
  Cohen, Huynh, Patterson, Yang, Taniguchi, Watanabe, and
  Young}}]{AndreaYoungBernal}
\bibinfo{author}{\bibfnamefont{H.}~\bibnamefont{Zhou}},
  \bibinfo{author}{\bibfnamefont{L.}~\bibnamefont{Holleis}},
  \bibinfo{author}{\bibfnamefont{Y.}~\bibnamefont{Saito}},
  \bibinfo{author}{\bibfnamefont{L.}~\bibnamefont{Cohen}},
  \bibinfo{author}{\bibfnamefont{W.}~\bibnamefont{Huynh}},
  \bibinfo{author}{\bibfnamefont{C.~L.} \bibnamefont{Patterson}},
  \bibinfo{author}{\bibfnamefont{F.}~\bibnamefont{Yang}},
  \bibinfo{author}{\bibfnamefont{T.}~\bibnamefont{Taniguchi}},
  \bibinfo{author}{\bibfnamefont{K.}~\bibnamefont{Watanabe}}, \bibnamefont{and}
  \bibinfo{author}{\bibfnamefont{A.~F.} \bibnamefont{Young}},
  \bibinfo{journal}{Science} \textbf{\bibinfo{volume}{375}},
  \bibinfo{pages}{774} (\bibinfo{year}{2022}),
  \urlprefix\url{https://doi.org/10.1126%2Fscience.abm8386}.

\bibitem[{\citenamefont{Zhou et~al.}(2021{\natexlab{a}})\citenamefont{Zhou,
  Xie, Taniguchi, Watanabe, and Young}}]{AndreaYoungRhombo}
\bibinfo{author}{\bibfnamefont{H.}~\bibnamefont{Zhou}},
  \bibinfo{author}{\bibfnamefont{T.}~\bibnamefont{Xie}},
  \bibinfo{author}{\bibfnamefont{T.}~\bibnamefont{Taniguchi}},
  \bibinfo{author}{\bibfnamefont{K.}~\bibnamefont{Watanabe}}, \bibnamefont{and}
  \bibinfo{author}{\bibfnamefont{A.~F.} \bibnamefont{Young}},
  \bibinfo{journal}{Nature} \textbf{\bibinfo{volume}{598}},
  \bibinfo{pages}{434} (\bibinfo{year}{2021}{\natexlab{a}}),
  \urlprefix\url{https://doi.org/10.1038%2Fs41586-021-03926-0}.

\bibitem[{\citenamefont{Zhou et~al.}(2021{\natexlab{b}})\citenamefont{Zhou,
  Xie, Ghazaryan, Holder, Ehrets, Spanton, Taniguchi, Watanabe, Berg, Serbyn
  et~al.}}]{AndreaYoungRhombo2}
\bibinfo{author}{\bibfnamefont{H.}~\bibnamefont{Zhou}},
  \bibinfo{author}{\bibfnamefont{T.}~\bibnamefont{Xie}},
  \bibinfo{author}{\bibfnamefont{A.}~\bibnamefont{Ghazaryan}},
  \bibinfo{author}{\bibfnamefont{T.}~\bibnamefont{Holder}},
  \bibinfo{author}{\bibfnamefont{J.~R.} \bibnamefont{Ehrets}},
  \bibinfo{author}{\bibfnamefont{E.~M.} \bibnamefont{Spanton}},
  \bibinfo{author}{\bibfnamefont{T.}~\bibnamefont{Taniguchi}},
  \bibinfo{author}{\bibfnamefont{K.}~\bibnamefont{Watanabe}},
  \bibinfo{author}{\bibfnamefont{E.}~\bibnamefont{Berg}},
  \bibinfo{author}{\bibfnamefont{M.}~\bibnamefont{Serbyn}},
  \bibnamefont{et~al.}, \bibinfo{journal}{Nature}
  \textbf{\bibinfo{volume}{598}}, \bibinfo{pages}{429}
  (\bibinfo{year}{2021}{\natexlab{b}}),
  \urlprefix\url{https://doi.org/10.1038%2Fs41586-021-03938-w}.

\bibitem[{\citenamefont{Serlin et~al.}(2020)\citenamefont{Serlin, Tschirhart,
  Polshyn, Zhang, Zhu, Watanabe, Taniguchi, Balents, and Young}}]{Serlin_2020}
\bibinfo{author}{\bibfnamefont{M.}~\bibnamefont{Serlin}},
  \bibinfo{author}{\bibfnamefont{C.~L.} \bibnamefont{Tschirhart}},
  \bibinfo{author}{\bibfnamefont{H.}~\bibnamefont{Polshyn}},
  \bibinfo{author}{\bibfnamefont{Y.}~\bibnamefont{Zhang}},
  \bibinfo{author}{\bibfnamefont{J.}~\bibnamefont{Zhu}},
  \bibinfo{author}{\bibfnamefont{K.}~\bibnamefont{Watanabe}},
  \bibinfo{author}{\bibfnamefont{T.}~\bibnamefont{Taniguchi}},
  \bibinfo{author}{\bibfnamefont{L.}~\bibnamefont{Balents}}, \bibnamefont{and}
  \bibinfo{author}{\bibfnamefont{A.~F.} \bibnamefont{Young}},
  \bibinfo{journal}{Science} \textbf{\bibinfo{volume}{367}},
  \bibinfo{pages}{900} (\bibinfo{year}{2020}),
  \urlprefix\url{https://doi.org/10.1126%2Fscience.aay5533}.

\bibitem[{\citenamefont{Wu et~al.}(2019{\natexlab{b}})\citenamefont{Wu, Lovorn,
  Tutuc, Martin, and MacDonald}}]{Wu_2019_TIPRL}
\bibinfo{author}{\bibfnamefont{F.}~\bibnamefont{Wu}},
  \bibinfo{author}{\bibfnamefont{T.}~\bibnamefont{Lovorn}},
  \bibinfo{author}{\bibfnamefont{E.}~\bibnamefont{Tutuc}},
  \bibinfo{author}{\bibfnamefont{I.}~\bibnamefont{Martin}}, \bibnamefont{and}
  \bibinfo{author}{\bibfnamefont{A.~H.} \bibnamefont{MacDonald}},
  \bibinfo{journal}{Phys. Rev. Lett.} \textbf{\bibinfo{volume}{122}},
  \bibinfo{pages}{086402} (\bibinfo{year}{2019}{\natexlab{b}}),
  \urlprefix\url{https://link.aps.org/doi/10.1103/PhysRevLett.122.086402}.

\bibitem[{\citenamefont{Tschirhart et~al.}(2022)\citenamefont{Tschirhart,
  Redekop, Li, Li, Jiang, Arp, Sheekey, Taniguchi, Watanabe, Mak
  et~al.}}]{KinFaiMak_TopologyTMD}
\bibinfo{author}{\bibfnamefont{C.~L.} \bibnamefont{Tschirhart}},
  \bibinfo{author}{\bibfnamefont{E.}~\bibnamefont{Redekop}},
  \bibinfo{author}{\bibfnamefont{L.}~\bibnamefont{Li}},
  \bibinfo{author}{\bibfnamefont{T.}~\bibnamefont{Li}},
  \bibinfo{author}{\bibfnamefont{S.}~\bibnamefont{Jiang}},
  \bibinfo{author}{\bibfnamefont{T.}~\bibnamefont{Arp}},
  \bibinfo{author}{\bibfnamefont{O.}~\bibnamefont{Sheekey}},
  \bibinfo{author}{\bibfnamefont{T.}~\bibnamefont{Taniguchi}},
  \bibinfo{author}{\bibfnamefont{K.}~\bibnamefont{Watanabe}},
  \bibinfo{author}{\bibfnamefont{K.~F.} \bibnamefont{Mak}},
  \bibnamefont{et~al.}, \emph{\bibinfo{title}{Intrinsic spin hall torque in a
  moire chern magnet}} (\bibinfo{year}{2022}),
  \urlprefix\url{https://arxiv.org/abs/2205.02823}.

\bibitem[{\citenamefont{Polshyn et~al.}(2020)\citenamefont{Polshyn, Zhu, Kumar,
  Zhang, Yang, Tschirhart, Serlin, Watanabe, Taniguchi, MacDonald
  et~al.}}]{Polshyn_2020}
\bibinfo{author}{\bibfnamefont{H.}~\bibnamefont{Polshyn}},
  \bibinfo{author}{\bibfnamefont{J.}~\bibnamefont{Zhu}},
  \bibinfo{author}{\bibfnamefont{M.~A.} \bibnamefont{Kumar}},
  \bibinfo{author}{\bibfnamefont{Y.}~\bibnamefont{Zhang}},
  \bibinfo{author}{\bibfnamefont{F.}~\bibnamefont{Yang}},
  \bibinfo{author}{\bibfnamefont{C.~L.} \bibnamefont{Tschirhart}},
  \bibinfo{author}{\bibfnamefont{M.}~\bibnamefont{Serlin}},
  \bibinfo{author}{\bibfnamefont{K.}~\bibnamefont{Watanabe}},
  \bibinfo{author}{\bibfnamefont{T.}~\bibnamefont{Taniguchi}},
  \bibinfo{author}{\bibfnamefont{A.~H.} \bibnamefont{MacDonald}},
  \bibnamefont{et~al.}, \bibinfo{journal}{Nature}
  \textbf{\bibinfo{volume}{588}}, \bibinfo{pages}{66} (\bibinfo{year}{2020}),
  \urlprefix\url{https://doi.org/10.1038%2Fs41586-020-2963-8}.

\bibitem[{\citenamefont{Zhang et~al.}(2022)\citenamefont{Zhang, Polski,
  Thomson, Lantagne-Hurtubise, Lewandowski, Zhou, Watanabe, Taniguchi, Alicea,
  and Nadj-Perge}}]{CalTechBernal}
\bibinfo{author}{\bibfnamefont{Y.}~\bibnamefont{Zhang}},
  \bibinfo{author}{\bibfnamefont{R.}~\bibnamefont{Polski}},
  \bibinfo{author}{\bibfnamefont{A.}~\bibnamefont{Thomson}},
  \bibinfo{author}{\bibfnamefont{Ã.}~\bibnamefont{Lantagne-Hurtubise}},
  \bibinfo{author}{\bibfnamefont{C.}~\bibnamefont{Lewandowski}},
  \bibinfo{author}{\bibfnamefont{H.}~\bibnamefont{Zhou}},
  \bibinfo{author}{\bibfnamefont{K.}~\bibnamefont{Watanabe}},
  \bibinfo{author}{\bibfnamefont{T.}~\bibnamefont{Taniguchi}},
  \bibinfo{author}{\bibfnamefont{J.}~\bibnamefont{Alicea}}, \bibnamefont{and}
  \bibinfo{author}{\bibfnamefont{S.}~\bibnamefont{Nadj-Perge}},
  \emph{\bibinfo{title}{Spin-orbit enhanced superconductivity in bernal bilayer
  graphene}} (\bibinfo{year}{2022}),
  \urlprefix\url{https://arxiv.org/abs/2205.05087}.

\bibitem[{\citenamefont{Polski et~al.}(2022)\citenamefont{Polski, Zhang, Peng,
  Arora, Choi, Kim, Watanabe, Taniguchi, Refael, von Oppen
  et~al.}}]{CalTechSymmetryBreakingTBLG}
\bibinfo{author}{\bibfnamefont{R.}~\bibnamefont{Polski}},
  \bibinfo{author}{\bibfnamefont{Y.}~\bibnamefont{Zhang}},
  \bibinfo{author}{\bibfnamefont{Y.}~\bibnamefont{Peng}},
  \bibinfo{author}{\bibfnamefont{H.~S.} \bibnamefont{Arora}},
  \bibinfo{author}{\bibfnamefont{Y.}~\bibnamefont{Choi}},
  \bibinfo{author}{\bibfnamefont{H.}~\bibnamefont{Kim}},
  \bibinfo{author}{\bibfnamefont{K.}~\bibnamefont{Watanabe}},
  \bibinfo{author}{\bibfnamefont{T.}~\bibnamefont{Taniguchi}},
  \bibinfo{author}{\bibfnamefont{G.}~\bibnamefont{Refael}},
  \bibinfo{author}{\bibfnamefont{F.}~\bibnamefont{von Oppen}},
  \bibnamefont{et~al.}, \emph{\bibinfo{title}{Hierarchy of symmetry breaking
  correlated phases in twisted bilayer graphene}} (\bibinfo{year}{2022}),
  \urlprefix\url{https://arxiv.org/abs/2205.05225}.

\bibitem[{\citenamefont{Xie and MacDonald}(2020)}]{Xie_2020}
\bibinfo{author}{\bibfnamefont{M.}~\bibnamefont{Xie}} \bibnamefont{and}
  \bibinfo{author}{\bibfnamefont{A.}~\bibnamefont{MacDonald}},
  \bibinfo{journal}{Physical Review Letters} \textbf{\bibinfo{volume}{124}}
  (\bibinfo{year}{2020}),
  \urlprefix\url{https://doi.org/10.1103%2Fphysrevlett.124.097601}.

\bibitem[{\citenamefont{Andrei and MacDonald}(2020)}]{MacDonaldTLBGReview}
\bibinfo{author}{\bibfnamefont{E.~Y.} \bibnamefont{Andrei}} \bibnamefont{and}
  \bibinfo{author}{\bibfnamefont{A.~H.} \bibnamefont{MacDonald}},
  \bibinfo{journal}{Nature Materials} \textbf{\bibinfo{volume}{19}},
  \bibinfo{pages}{1265} (\bibinfo{year}{2020}),
  \urlprefix\url{https://doi.org/10.1038%2Fs41563-020-00840-0}.

\bibitem[{\citenamefont{Li et~al.}(2021)\citenamefont{Li, Jiang, Li, Zhang,
  Kang, Zhu, Watanabe, Taniguchi, Chowdhury, Fu et~al.}}]{Li_2021}
\bibinfo{author}{\bibfnamefont{T.}~\bibnamefont{Li}},
  \bibinfo{author}{\bibfnamefont{S.}~\bibnamefont{Jiang}},
  \bibinfo{author}{\bibfnamefont{L.}~\bibnamefont{Li}},
  \bibinfo{author}{\bibfnamefont{Y.}~\bibnamefont{Zhang}},
  \bibinfo{author}{\bibfnamefont{K.}~\bibnamefont{Kang}},
  \bibinfo{author}{\bibfnamefont{J.}~\bibnamefont{Zhu}},
  \bibinfo{author}{\bibfnamefont{K.}~\bibnamefont{Watanabe}},
  \bibinfo{author}{\bibfnamefont{T.}~\bibnamefont{Taniguchi}},
  \bibinfo{author}{\bibfnamefont{D.}~\bibnamefont{Chowdhury}},
  \bibinfo{author}{\bibfnamefont{L.}~\bibnamefont{Fu}}, \bibnamefont{et~al.},
  \bibinfo{journal}{Nature} \textbf{\bibinfo{volume}{597}},
  \bibinfo{pages}{350} (\bibinfo{year}{2021}),
  \urlprefix\url{https://doi.org/10.1038%2Fs41586-021-03853-0}.

\bibitem[{\citenamefont{Ghiotto et~al.}(2021)\citenamefont{Ghiotto, Shih,
  Pereira, Rhodes, Kim, Zang, Millis, Watanabe, Taniguchi, Hone
  et~al.}}]{Ghiotto_2021}
\bibinfo{author}{\bibfnamefont{A.}~\bibnamefont{Ghiotto}},
  \bibinfo{author}{\bibfnamefont{E.-M.} \bibnamefont{Shih}},
  \bibinfo{author}{\bibfnamefont{G.~S. S.~G.} \bibnamefont{Pereira}},
  \bibinfo{author}{\bibfnamefont{D.~A.} \bibnamefont{Rhodes}},
  \bibinfo{author}{\bibfnamefont{B.}~\bibnamefont{Kim}},
  \bibinfo{author}{\bibfnamefont{J.}~\bibnamefont{Zang}},
  \bibinfo{author}{\bibfnamefont{A.~J.} \bibnamefont{Millis}},
  \bibinfo{author}{\bibfnamefont{K.}~\bibnamefont{Watanabe}},
  \bibinfo{author}{\bibfnamefont{T.}~\bibnamefont{Taniguchi}},
  \bibinfo{author}{\bibfnamefont{J.~C.} \bibnamefont{Hone}},
  \bibnamefont{et~al.}, \bibinfo{journal}{Nature}
  \textbf{\bibinfo{volume}{597}}, \bibinfo{pages}{345} (\bibinfo{year}{2021}),
  \urlprefix\url{https://doi.org/10.1038%2Fs41586-021-03815-6}.

\bibitem[{\citenamefont{Pan et~al.}(2020)\citenamefont{Pan, Wu, and
  Sarma}}]{Pan_2020}
\bibinfo{author}{\bibfnamefont{H.}~\bibnamefont{Pan}},
  \bibinfo{author}{\bibfnamefont{F.}~\bibnamefont{Wu}}, \bibnamefont{and}
  \bibinfo{author}{\bibfnamefont{S.~D.} \bibnamefont{Sarma}},
  \bibinfo{journal}{Physical Review B} \textbf{\bibinfo{volume}{102}}
  (\bibinfo{year}{2020}),
  \urlprefix\url{https://doi.org/10.1103%2Fphysrevb.102.201104}.

\bibitem[{\citenamefont{Pan and Sarma}(2021)}]{Pan_2021}
\bibinfo{author}{\bibfnamefont{H.}~\bibnamefont{Pan}} \bibnamefont{and}
  \bibinfo{author}{\bibfnamefont{S.~D.} \bibnamefont{Sarma}},
  \bibinfo{journal}{Physical Review Letters} \textbf{\bibinfo{volume}{127}}
  (\bibinfo{year}{2021}),
  \urlprefix\url{https://doi.org/10.1103%2Fphysrevlett.127.096802}.

\bibitem[{\citenamefont{Morales-Dur{\'{a}}n
  et~al.}(2021)\citenamefont{Morales-Dur{\'{a}}n, MacDonald, and
  Potasz}}]{Morales_Dur_n_2021}
\bibinfo{author}{\bibfnamefont{N.}~\bibnamefont{Morales-Dur{\'{a}}n}},
  \bibinfo{author}{\bibfnamefont{A.~H.} \bibnamefont{MacDonald}},
  \bibnamefont{and} \bibinfo{author}{\bibfnamefont{P.}~\bibnamefont{Potasz}},
  \bibinfo{journal}{Physical Review B} \textbf{\bibinfo{volume}{103}}
  (\bibinfo{year}{2021}),
  \urlprefix\url{https://doi.org/10.1103%2Fphysrevb.103.l241110}.

\bibitem[{\citenamefont{Ahn and Sarma}(2022)}]{Ahn_2022}
\bibinfo{author}{\bibfnamefont{S.}~\bibnamefont{Ahn}} \bibnamefont{and}
  \bibinfo{author}{\bibfnamefont{S.~D.} \bibnamefont{Sarma}},
  \bibinfo{journal}{Physical Review B} \textbf{\bibinfo{volume}{105}}
  (\bibinfo{year}{2022}),
  \urlprefix\url{https://doi.org/10.1103%2Fphysrevb.105.115114}.

\bibitem[{\citenamefont{Kerelsky et~al.}(2021)\citenamefont{Kerelsky,
  Rubio-Verd{\'u}, Xian, Kennes, Halbertal, Finney, Song, Turkel, Wang,
  Watanabe et~al.}}]{Kerelsky2021moireless}
\bibinfo{author}{\bibfnamefont{A.}~\bibnamefont{Kerelsky}},
  \bibinfo{author}{\bibfnamefont{C.}~\bibnamefont{Rubio-Verd{\'u}}},
  \bibinfo{author}{\bibfnamefont{L.}~\bibnamefont{Xian}},
  \bibinfo{author}{\bibfnamefont{D.~M.} \bibnamefont{Kennes}},
  \bibinfo{author}{\bibfnamefont{D.}~\bibnamefont{Halbertal}},
  \bibinfo{author}{\bibfnamefont{N.}~\bibnamefont{Finney}},
  \bibinfo{author}{\bibfnamefont{L.}~\bibnamefont{Song}},
  \bibinfo{author}{\bibfnamefont{S.}~\bibnamefont{Turkel}},
  \bibinfo{author}{\bibfnamefont{L.}~\bibnamefont{Wang}},
  \bibinfo{author}{\bibfnamefont{K.}~\bibnamefont{Watanabe}},
  \bibnamefont{et~al.}, \bibinfo{journal}{Proceedings of the National Academy
  of Sciences} \textbf{\bibinfo{volume}{118}} (\bibinfo{year}{2021}).

\bibitem[{\citenamefont{Khalaf et~al.}(2019)\citenamefont{Khalaf, Kruchkov,
  Tarnopolsky, and Vishwanath}}]{Khalaf2019}
\bibinfo{author}{\bibfnamefont{E.}~\bibnamefont{Khalaf}},
  \bibinfo{author}{\bibfnamefont{A.~J.} \bibnamefont{Kruchkov}},
  \bibinfo{author}{\bibfnamefont{G.}~\bibnamefont{Tarnopolsky}},
  \bibnamefont{and}
  \bibinfo{author}{\bibfnamefont{A.}~\bibnamefont{Vishwanath}},
  \bibinfo{journal}{Phys. Rev. B} \textbf{\bibinfo{volume}{100}},
  \bibinfo{pages}{085109} (\bibinfo{year}{2019}),
  \urlprefix\url{https://link.aps.org/doi/10.1103/PhysRevB.100.085109}.

\bibitem[{\citenamefont{Arora et~al.}(2020)\citenamefont{Arora, Polski, Zhang,
  Thomson, Choi, Kim, Lin, Wilson, Xu, Chu et~al.}}]{CalTechTwisted}
\bibinfo{author}{\bibfnamefont{H.~S.} \bibnamefont{Arora}},
  \bibinfo{author}{\bibfnamefont{R.}~\bibnamefont{Polski}},
  \bibinfo{author}{\bibfnamefont{Y.}~\bibnamefont{Zhang}},
  \bibinfo{author}{\bibfnamefont{A.}~\bibnamefont{Thomson}},
  \bibinfo{author}{\bibfnamefont{Y.}~\bibnamefont{Choi}},
  \bibinfo{author}{\bibfnamefont{H.}~\bibnamefont{Kim}},
  \bibinfo{author}{\bibfnamefont{Z.}~\bibnamefont{Lin}},
  \bibinfo{author}{\bibfnamefont{I.~Z.} \bibnamefont{Wilson}},
  \bibinfo{author}{\bibfnamefont{X.}~\bibnamefont{Xu}},
  \bibinfo{author}{\bibfnamefont{J.-H.} \bibnamefont{Chu}},
  \bibnamefont{et~al.}, \bibinfo{journal}{Nature}
  \textbf{\bibinfo{volume}{583}}, \bibinfo{pages}{379} (\bibinfo{year}{2020}),
  \urlprefix\url{https://doi.org/10.1038%2Fs41586-020-2473-8}.

\bibitem[{\citenamefont{Ahn and Das~Sarma}(2022)}]{Ahn_2022_Anderson}
\bibinfo{author}{\bibfnamefont{S.}~\bibnamefont{Ahn}} \bibnamefont{and}
  \bibinfo{author}{\bibfnamefont{S.}~\bibnamefont{Das~Sarma}},
  \bibinfo{journal}{Phys. Rev. Mater.} \textbf{\bibinfo{volume}{6}},
  \bibinfo{pages}{L091001} (\bibinfo{year}{2022}),
  \urlprefix\url{https://link.aps.org/doi/10.1103/PhysRevMaterials.6.L091001}.

\bibitem[{\citenamefont{Davis et~al.}(2023)\citenamefont{Davis, Chou, Wu, and
  Das~Sarma}}]{Davis_2022}
\bibinfo{author}{\bibfnamefont{S.~M.} \bibnamefont{Davis}},
  \bibinfo{author}{\bibfnamefont{Y.-Z.} \bibnamefont{Chou}},
  \bibinfo{author}{\bibfnamefont{F.}~\bibnamefont{Wu}}, \bibnamefont{and}
  \bibinfo{author}{\bibfnamefont{S.}~\bibnamefont{Das~Sarma}},
  \bibinfo{journal}{Phys. Rev. B} \textbf{\bibinfo{volume}{107}},
  \bibinfo{pages}{045426} (\bibinfo{year}{2023}),
  \urlprefix\url{https://link.aps.org/doi/10.1103/PhysRevB.107.045426}.

\bibitem[{\citenamefont{Coleman}(2015)}]{Coleman2015introduction}
\bibinfo{author}{\bibfnamefont{P.}~\bibnamefont{Coleman}},
  \emph{\bibinfo{title}{Introduction to Many-Body Physics}}
  (\bibinfo{publisher}{Cambridge University Press}, \bibinfo{year}{2015}), ISBN
  \bibinfo{isbn}{9780521864886}.

\bibitem[{\citenamefont{Jaoui et~al.}(2022)\citenamefont{Jaoui, Das, Battista,
  D{\'{\i}}ez-M{\'{e}}rida, Lu, Watanabe, Taniguchi, Ishizuka, Levitov, and
  Efetov}}]{Jaoui_2022}
\bibinfo{author}{\bibfnamefont{A.}~\bibnamefont{Jaoui}},
  \bibinfo{author}{\bibfnamefont{I.}~\bibnamefont{Das}},
  \bibinfo{author}{\bibfnamefont{G.~D.} \bibnamefont{Battista}},
  \bibinfo{author}{\bibfnamefont{J.}~\bibnamefont{D{\'{\i}}ez-M{\'{e}}rida}},
  \bibinfo{author}{\bibfnamefont{X.}~\bibnamefont{Lu}},
  \bibinfo{author}{\bibfnamefont{K.}~\bibnamefont{Watanabe}},
  \bibinfo{author}{\bibfnamefont{T.}~\bibnamefont{Taniguchi}},
  \bibinfo{author}{\bibfnamefont{H.}~\bibnamefont{Ishizuka}},
  \bibinfo{author}{\bibfnamefont{L.}~\bibnamefont{Levitov}}, \bibnamefont{and}
  \bibinfo{author}{\bibfnamefont{D.~K.} \bibnamefont{Efetov}},
  \bibinfo{journal}{Nature Physics} \textbf{\bibinfo{volume}{18}},
  \bibinfo{pages}{633} (\bibinfo{year}{2022}),
  \urlprefix\url{https://doi.org/10.1038%2Fs41567-022-01556-5}.

\bibitem[{\citenamefont{Davis and Sarma}(2023)}]{Davis_2023Subsonic}
\bibinfo{author}{\bibfnamefont{S.~M.} \bibnamefont{Davis}} \bibnamefont{and}
  \bibinfo{author}{\bibfnamefont{S.~D.} \bibnamefont{Sarma}},
  \emph{\bibinfo{title}{The kinetic theory of ultra-subsonic fermion systems
  and applications to flat band magic angle twisted bilayer graphene}}
  (\bibinfo{year}{2023}), \eprint{2305.09120}.

\bibitem[{\citenamefont{Prange and Kadanoff}(1964)}]{Prange_1964}
\bibinfo{author}{\bibfnamefont{R.~E.} \bibnamefont{Prange}} \bibnamefont{and}
  \bibinfo{author}{\bibfnamefont{L.~P.} \bibnamefont{Kadanoff}},
  \bibinfo{journal}{Phys. Rev.} \textbf{\bibinfo{volume}{134}},
  \bibinfo{pages}{A566} (\bibinfo{year}{1964}),
  \urlprefix\url{https://link.aps.org/doi/10.1103/PhysRev.134.A566}.

\bibitem[{\citenamefont{Prange and Sachs}(1967)}]{Prange_1967}
\bibinfo{author}{\bibfnamefont{R.~E.} \bibnamefont{Prange}} \bibnamefont{and}
  \bibinfo{author}{\bibfnamefont{A.}~\bibnamefont{Sachs}},
  \bibinfo{journal}{Phys. Rev.} \textbf{\bibinfo{volume}{158}},
  \bibinfo{pages}{672} (\bibinfo{year}{1967}),
  \urlprefix\url{https://link.aps.org/doi/10.1103/PhysRev.158.672}.

\bibitem[{\citenamefont{Choi et~al.}(2021)\citenamefont{Choi, Kim, Lewandowski,
  Peng, Thomson, Polski, Zhang, Watanabe, Taniguchi, Alicea
  et~al.}}]{Choi_2021_InteractionDriven}
\bibinfo{author}{\bibfnamefont{Y.}~\bibnamefont{Choi}},
  \bibinfo{author}{\bibfnamefont{H.}~\bibnamefont{Kim}},
  \bibinfo{author}{\bibfnamefont{C.}~\bibnamefont{Lewandowski}},
  \bibinfo{author}{\bibfnamefont{Y.}~\bibnamefont{Peng}},
  \bibinfo{author}{\bibfnamefont{A.}~\bibnamefont{Thomson}},
  \bibinfo{author}{\bibfnamefont{R.}~\bibnamefont{Polski}},
  \bibinfo{author}{\bibfnamefont{Y.}~\bibnamefont{Zhang}},
  \bibinfo{author}{\bibfnamefont{K.}~\bibnamefont{Watanabe}},
  \bibinfo{author}{\bibfnamefont{T.}~\bibnamefont{Taniguchi}},
  \bibinfo{author}{\bibfnamefont{J.}~\bibnamefont{Alicea}},
  \bibnamefont{et~al.}, \bibinfo{journal}{Nature Physics}
  \textbf{\bibinfo{volume}{17}} (\bibinfo{year}{2021}),
  \urlprefix\url{https://doi.org/10.1038/s41567-021-01359-0}.

\bibitem[{\citenamefont{Lewandowski et~al.}(2021)\citenamefont{Lewandowski,
  Nadj-Perge, and Chowdhury}}]{Lewandowski_2021_FillingDependent}
\bibinfo{author}{\bibfnamefont{C.}~\bibnamefont{Lewandowski}},
  \bibinfo{author}{\bibfnamefont{S.}~\bibnamefont{Nadj-Perge}},
  \bibnamefont{and}
  \bibinfo{author}{\bibfnamefont{D.}~\bibnamefont{Chowdhury}},
  \bibinfo{journal}{npj Quantum Materials} \textbf{\bibinfo{volume}{6}}
  (\bibinfo{year}{2021}),
  \urlprefix\url{https://doi.org/10.1038%2Fs41535-021-00379-6}.

\bibitem[{\citenamefont{Rademaker et~al.}(2019)\citenamefont{Rademaker, Abanin,
  and Mellado}}]{Rademaker_2019}
\bibinfo{author}{\bibfnamefont{L.}~\bibnamefont{Rademaker}},
  \bibinfo{author}{\bibfnamefont{D.~A.} \bibnamefont{Abanin}},
  \bibnamefont{and} \bibinfo{author}{\bibfnamefont{P.}~\bibnamefont{Mellado}},
  \bibinfo{journal}{Phys. Rev. B} \textbf{\bibinfo{volume}{100}},
  \bibinfo{pages}{205114} (\bibinfo{year}{2019}),
  \urlprefix\url{https://link.aps.org/doi/10.1103/PhysRevB.100.205114}.

\bibitem[{\citenamefont{Wong et~al.}(2020)\citenamefont{Wong, Nuckolls, Oh,
  Lian, Xie, Jeon, Watanabe, Taniguchi, Bernevig, and Yazdani}}]{Wong_2020}
\bibinfo{author}{\bibfnamefont{D.}~\bibnamefont{Wong}},
  \bibinfo{author}{\bibfnamefont{K.~P.} \bibnamefont{Nuckolls}},
  \bibinfo{author}{\bibfnamefont{M.}~\bibnamefont{Oh}},
  \bibinfo{author}{\bibfnamefont{B.}~\bibnamefont{Lian}},
  \bibinfo{author}{\bibfnamefont{Y.}~\bibnamefont{Xie}},
  \bibinfo{author}{\bibfnamefont{S.}~\bibnamefont{Jeon}},
  \bibinfo{author}{\bibfnamefont{K.}~\bibnamefont{Watanabe}},
  \bibinfo{author}{\bibfnamefont{T.}~\bibnamefont{Taniguchi}},
  \bibinfo{author}{\bibfnamefont{B.~A.} \bibnamefont{Bernevig}},
  \bibnamefont{and} \bibinfo{author}{\bibfnamefont{A.}~\bibnamefont{Yazdani}},
  \bibinfo{journal}{Nature} \textbf{\bibinfo{volume}{582}},
  \bibinfo{pages}{198} (\bibinfo{year}{2020}),
  \urlprefix\url{https://doi.org/10.1038%2Fs41586-020-2339-0}.

\bibitem[{\citenamefont{Zondiner et~al.}(2020)\citenamefont{Zondiner, Rozen,
  Rodan-Legrain, Cao, Queiroz, Taniguchi, Watanabe, Oreg, von Oppen, Stern
  et~al.}}]{Zondiner_2020}
\bibinfo{author}{\bibfnamefont{U.}~\bibnamefont{Zondiner}},
  \bibinfo{author}{\bibfnamefont{A.}~\bibnamefont{Rozen}},
  \bibinfo{author}{\bibfnamefont{D.}~\bibnamefont{Rodan-Legrain}},
  \bibinfo{author}{\bibfnamefont{Y.}~\bibnamefont{Cao}},
  \bibinfo{author}{\bibfnamefont{R.}~\bibnamefont{Queiroz}},
  \bibinfo{author}{\bibfnamefont{T.}~\bibnamefont{Taniguchi}},
  \bibinfo{author}{\bibfnamefont{K.}~\bibnamefont{Watanabe}},
  \bibinfo{author}{\bibfnamefont{Y.}~\bibnamefont{Oreg}},
  \bibinfo{author}{\bibfnamefont{F.}~\bibnamefont{von Oppen}},
  \bibinfo{author}{\bibfnamefont{A.}~\bibnamefont{Stern}},
  \bibnamefont{et~al.}, \bibinfo{journal}{Nature}
  \textbf{\bibinfo{volume}{582}}, \bibinfo{pages}{203} (\bibinfo{year}{2020}),
  \urlprefix\url{https://doi.org/10.1038%2Fs41586-020-2373-y}.

\end{thebibliography}

\end{document}